%% file: unsteady.tex
\documentclass[pre,notitlepage,tightenlines,nofootinbib,12pt]{revtex4-1}

\usepackage{amsmath,amssymb,amsfonts}
\usepackage{bm}
\usepackage{enumerate}
\usepackage{hyperref}
\usepackage{color}
\usepackage{graphicx}
\usepackage{caption}
\usepackage{subcaption}
\usepackage{tikz}
\usepackage{mathtools}

\graphicspath{{figs/}}

\DeclareMathOperator{\erf}{erf}

\newcommand{\tavg}[1]{\langle #1 \rangle}
\newcommand{\Order}[1]{O(#1)}
\newcommand{\mathnotation}[2]{\newcommand{#1}{\ensuremath{#2}}}

\renewcommand{\l}{\left}
\renewcommand{\r}{\right}
\renewcommand{\time}{t}                
\mathnotation{\stime}{s}               
\mathnotation{\Time}{T}                
\mathnotation{\pd}{\partial}           
\mathnotation{\imi}{\mathrm{i}}        
\mathnotation{\dint}{\,{\mathrm{d}}}   
\mathnotation{\ldef}{\mathrel{\raisebox{.069ex}{:}\!\!=}}
\mathnotation{\rdef}{\mathrel{=\!\!\raisebox{.069ex}{:}}}
\mathnotation{\ee}{\mathrm{e}}         
\mathnotation{\eps}{\varepsilon}       

\mathnotation{\grad}{\nabla}
\mathnotation{\lapl}{\grad^2}
\renewcommand{\div}{\grad\cdot}

\mathnotation{\rc}{r}
\mathnotation{\rv}{\bm{\rc}}
\mathnotation{\rvp}{\rv}
\mathnotation{\ruv}{\hat{\bm{\rc}}}
\mathnotation{\xc}{x}
\mathnotation{\xcp}{\xc}
\mathnotation{\yc}{y}
\mathnotation{\ycp}{\yc}
\mathnotation{\yv}{\bm{\yc}}
\mathnotation{\zc}{z}
\mathnotation{\zcp}{\zc}
\mathnotation{\xuv}{\hat{\bm{\xc}}}
\mathnotation{\yuv}{\hat{\bm{\yc}}}
\mathnotation{\zuv}{\hat{\bm{\zc}}}

\mathnotation{\uc}{u}                       
\mathnotation{\uv}{\bm{u}}                  
\mathnotation{\uvstress}{\bm{\nu}_{\text{stress}}}
\mathnotation{\id}{\mathbb{I}}              
\mathnotation{\Fv}{\bm{F}}                  
\mathnotation{\fv}{\bm{f}}                  
\mathnotation{\F}{F}                        
\mathnotation{\f}{f}                        
\mathnotation{\Os}{\mathbb{G}}              

\mathnotation{\Uc}{U}                       
\mathnotation{\Uv}{\bm{\Uc}}                
\mathnotation{\RR}{R}                       
\mathnotation{\rr}{\rho}                    
\renewcommand{\aa}{a}                       
\mathnotation{\da}{\skew{3}\dot{\aa}}       
\renewcommand{\AA}{A}                       
\mathnotation{\dA}{\skew{6}{\dot}{\AA}}     
\mathnotation{\Asep}{\alpha}                
\mathnotation{\Asepv}{\bm{\Asep}}           
\mathnotation{\RRAsep}{c_{\Asep}}           
\mathnotation{\StImagev}{\Asepv^*}          
\mathnotation{\Hyp}{d}                      
\mathnotation{\II}{I}                       
\mathnotation{\Deltafar}{\Delta}
\mathnotation{\ndens}{n}                    
\mathnotation{\pres}{p}                     
\mathnotation{\W}{W}                        
\mathnotation{\Wavg}{\mathcal{W}}           
\mathnotation{\Tres}{T_{\text{res}}}        
\mathnotation{\Cnear}{C}                    
\newcommand{\betak}[1]{\beta_{(#1)}}        
\newcommand{\rvk}[1]{\rv_{(#1)}}            

\mathnotation{\microm}{\mu\mathrm{m}}
\mathnotation{\second}{\mathrm{s}}
\mathnotation{\Hertz}{\mathrm{Hz}}
\mathnotation{\radian}{\mathrm{rad}}
\mathnotation{\lengthscale}{R}
\mathnotation{\timescale}{T}

\mathnotation{\Deffective}{D_{\mathrm{eff}}}
\mathnotation{\Dthermal}{D_0}               
\mathnotation{\Denhanced}{D_\mathrm{h}}     
\newcommand{\enhanced}{enhanced}            

\begin{document}

\title{Fluid transport and mixing by an unsteady microswimmer}
\author{Peter Mueller}
\author{Jean-Luc Thiffeault}
\affiliation{Department of Mathematics, University of Wisconsin --
  Madison, 480 Lincoln Dr., Madison, WI 53706, USA}

\begin{abstract}
  We study the fluid drift due to a time-dependent dumbbell model of a
  microswimmer. The model captures important aspects of real microswimmers
  such as a time-dependent flagellar motion and a no-slip body.  The model
  consists of a rigid sphere for the body and a time-dependent moving
  Stokeslet representing the flagella.  We analyze the paths of idealized
  fluid particles displaced by the swimmer.  The simplicity of the model
  allows some asymptotic calculations very near and far away from the swimmer.
  The displacements of particles near the swimmer diverge in a manner similar
  to an isolated no-slip sphere, but with a smaller coefficient due to the
  action of the flagellum.  Far from the swimmer, the time dependence becomes
  negligible due to both being very fast and decaying with distance.  Finally,
  we compute the probability distribution of particle displacements, and find
  that our model has fatter tails than previous steady models, due to the
  presence of a no-slip surface that drags particles along.
\end{abstract}



\maketitle

\section{Introduction}

Most microorganisms depend on a well-mixed environment for their supply of nutrients.  The nutrients typically have very slow rates of diffusion, so some amount of mechanical stirring is needed to enhance mixing.  This stirring is often caused by external factors such as winds, tides, or gravity waves in the ocean, or circulation in blood vessels.  However, the motion of the organisms themselves can assist this process.  The stirring and mixing of an environment caused by swimming organisms is called \emph{biogenic mixing}, or \emph{biomixing} for short.

Biomixing has been investigated for several years with the aid of experimental observations~\cite{Kim2004,Kunze2006,Chen2007,Katija2009,Leptos2009,Lorke2010,Kurtuldu2011,Katija2012,Jepson2013,Noss2014} as well as theoretical models and numerical simulations~\cite{Underhill2008,Rushkin2010,Ishikawa2010,Leshansky2010,Kunze2011,Lin2011,Zaid2011}. The importance of biomixing remains unclear in relation to mixing caused by winds, waves, molecular diffusion, and other factors~\cite{Dabiri2010,Dewar2006,Gregg2009,Huntley2004,Rousseau2010,Subramanian2010,Thiffeault2010b,Visser2007}.  Moreover, there are applications such as aquaculture where the density of swimmers can be controlled, and mixing is of crucial importance to the well-being of the organisms~\cite{Rasmussen2005}.  It is thus important to understand the detailed manner in which biomixing arises in order to gauge its possible impact.

At higher Reynolds numbers (when inertial effects are larger than viscous damping), mixing can be assisted by turbulence~\cite{Lorke2010}. The fluid motion due to microscopic swimmers (or microswimmers) normally has a very small Reynolds number. In this regime (known as Stokes flow), viscous dissipation dominates over inertial effects, and the Scallop theorem~\cite{Purcell1977} applies: a swimmer needs to make time-irreversible motions to make any progress.  Locomotion in the Stokes regime thus requires a carefully tailored approach.  The flows set up by the microswimmers decay slowly with distance, which enhances the diffusion of tracers such as nutrients. This is amplified by the abundance of microorganisms in the medium.

\begin{figure}
\begin{center}
\begin{subfigure}[b]{.4\textwidth}
  \input{figs_tikz/chlamy.tex}
  \caption{}
  \label{fig:chlamy}
\end{subfigure}
~
\begin{subfigure}[b]{.4\textwidth}
  \raisebox{.55in}{\input{figs_tikz/swimmer.tex}}
  \caption{}
  \label{fig:swimmer}
\end{subfigure}
\end{center}
\caption{(a) Diagram of \emph{Chlamydomonas reinhardtii} and the swimming
  stroke of its flagella. (b) The time-dependent dumbbell swimmer model,
  swimming to the right.}
\end{figure}
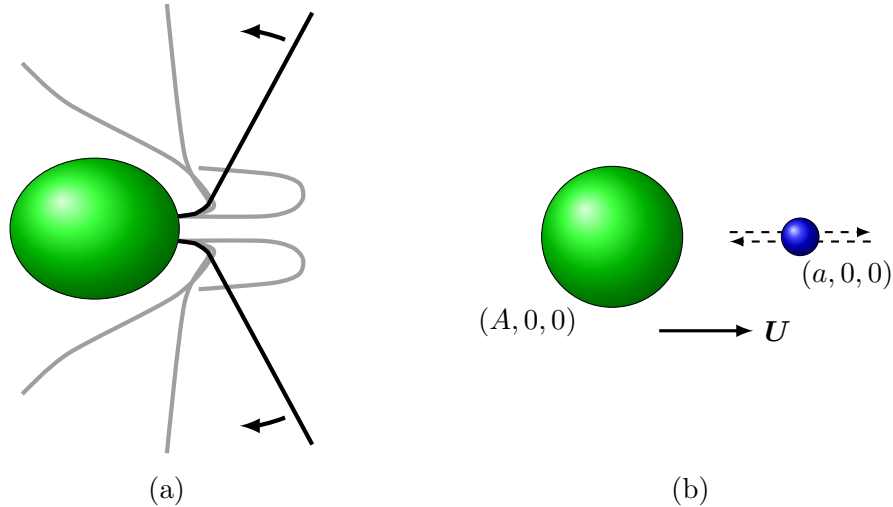

Microswimmers are generally grouped into two categories --- pushers and pullers --- based on the positioning of their propulsion mechanism. \emph{Escherichia coli} (a pusher) has a rotating helical filament located on its posterior end, while~\emph{Chlamydomonas reinhardtii} (a puller) has a pair of anterior flagella that move similar to that of a breast-stroke (see Fig.~\ref{fig:chlamy}). At high volume fractions of swimmers, pushers align with each other, creating mixing effects at larger scales~\cite{Mendelson1999,Dombrowski2004,HernandezOrtiz2005,Saintillan2007,Sokolov2007,Sokolov2009,Cisneros2011,Underhill2011,Saintillan2012}. Additionally, the same dynamics that cause the alignment of pushers also lead to attractions with surfaces as seen in~\cite{Rothschild1963,Winet1984,Cosson2003,Lauga2006,Berke2008,Drescher2009}.

Recent experiments have shown enhanced tracer diffusion at low volume fractions for pushers, pullers, and self-propelled particles~\cite{Wu2000,Leptos2009,Sokolov2009,Kurtuldu2011,Mino2011,Mino2013,Jepson2013}. Both simulations and theoretical arguments~\cite{Underhill2008,Ishikawa2010,Rushkin2010,Lin2011,Zaid2011,Pushkin2013} support this finding. Particles often traverse loop-like trajectories when a swimmer moves by, resulting in reduced net-displacements~\cite{Maxwell1869,Dunkel2010}. These closed trajectories can be opened up by stagnation points near the swimmer~\cite{Lin2011} and finite swimming paths~\cite{Thiffeault2010b,Pushkin2013}.  Some microorganisms exhibiting run-and-tumble behavior have natural finite path lengths; for example, \emph{E.\ coli} does this to traverse biochemical gradients. Other swimmers experience rotational diffusion or other environmental effects.

In order to explain enhanced tracer diffusion, the \emph{drift} caused by a swimmer must be analyzed.  Drift due to moving bodies is an interesting topic of study in its own right, and has been examined by \citet{Maxwell1869}, \citet{Darwin1953}, and \citet{Lighthill1956}.  The drift due to an isolated no-slip sphere in Stokes flow has been investigated by many, for instance by \citet{Eames2003b}.  Recently, there has been interest in drift due to wakes \cite{Lin2011,Melkoumian2014}, multiple objects \cite{Lin2016,Melkoumian2016}, and steady microswimmers~\cite{Dunkel2010,Thiffeault2010b,Lin2011,Pushkin2013}.  In the context of mixing by swimming organisms, \citet{Katija2009} and \citet{Thiffeault2010b} proposed that the \enhanced~diffusivity is the result of fluid particles interacting with many swimmers, and thus experiencing multiple drifts. \citet{Leptos2009} highlight that the unsteadiness of the flow contributes to the complex dynamics.

The majority of papers on swimmer suspensions use flows that are steady in the frame of the swimmers to simplify the problem.  (There are many exceptions, such as~\cite{Michelin2013}.)  The present paper explores the effect of time-dependence on the displacement of tracer particles. We then use the probabilistic model formulated by \citet{Thiffeault2010b} and \citet{Lin2011} to analyze the fluid mixing due to a collection of unsteady swimmers. This model has been recently tested in numerical simulations~\cite{Morozov2014,Kasyap2014} and shown to hold in more complex setups~\cite{Pushkin2013b,Pushkin2014}.

The outline of the rest of this paper is as follows.  In Section~\ref{sec:model} we describe a simple time-dependent model of a puller.  For simplicity, our swimmer has axial symmetry along the swimming axis which reduces the dimensionality of the problem. We compare the flow field of our model to recent experimental measurements of~\citet{Drescher2010} and~\citet{Guasto2010}. Then in Section~\ref{sec:numerical} we carry out numerical integration of particle trajectories for the time-dependent model.  Sections~\ref{sec:nearfield} and~\ref{sec:farfield} focus on asymptotic analysis of particle displacements near and far from the swimmer, respectively.

In Section~\ref{sec:stats}, we quantify the statistics of particle displacements due to a suspension of microswimmers.  We do this in two ways: first we evaluate the effective diffusivity imparted by the swimmers, then we compute the full probability distribution of particle displacements.  Finally, we offer some conclusions in Section~\ref{sec:conc}.


\section{Dumbbell swimmer model}
\label{sec:model}

\emph{Chlamydomonas reinhardtii} has a roughly spherical body and a pair of anterior flagella that it uses for locomotion (see Fig.~\ref{fig:chlamy}). \emph{C.\ reinhardtii}'s size ($4\,\microm$ radius) and speed ($100\,\microm/\second$) give a Reynolds number of about $10^{-4}$ in water. Even with a high beat frequency of~$50\,\Hertz$, the Strouhal number (a ratio of time scales involved with swimming to that of flagellar oscillations) is only 2, which means the steady Stokes equations accurately model fluid flow.  The asymmetric motion of \emph{C.\ reinhardtii}'s flagella enables it to swim in the Stokes regime.  In~\citet{Friedrich2012}, \emph{C.\ reinhardtii} is modeled with a sphere representing the body and two spheres for the flagella.  In their model the organism makes forward progress due to the asymmetric interactions between the two flagellar spheres during the power and recovery strokes.  Here we use a further simplification of this type of model, a time-dependent dumbbell.

\subsection{A time-dependent dumbbell model}
\label{subsec:model}
\label{subsec:paramsetderivation}

Our simplified model (pictured in Fig.~\ref{fig:swimmer}) involves only two spheres, one of which will be represented as a point force.  This gives the swimmer axial symmetry along its swimming direction, which will greatly facilitate numerical volume integration later. We approximate the body by a rigid sphere and the net propulsion of the flagella by a single Stokeslet, which allows an analytic solution. To achieve locomotion we allow the strength of the flagellar Stokeslet to vary. This represents the asymmetric drag due to the varying geometry of the flagella in the power and recovery strokes.

\begin{figure}
  \begin{center}
    \begin{subfigure}[b]{.24\textwidth}
      \includegraphics[width=\textwidth]{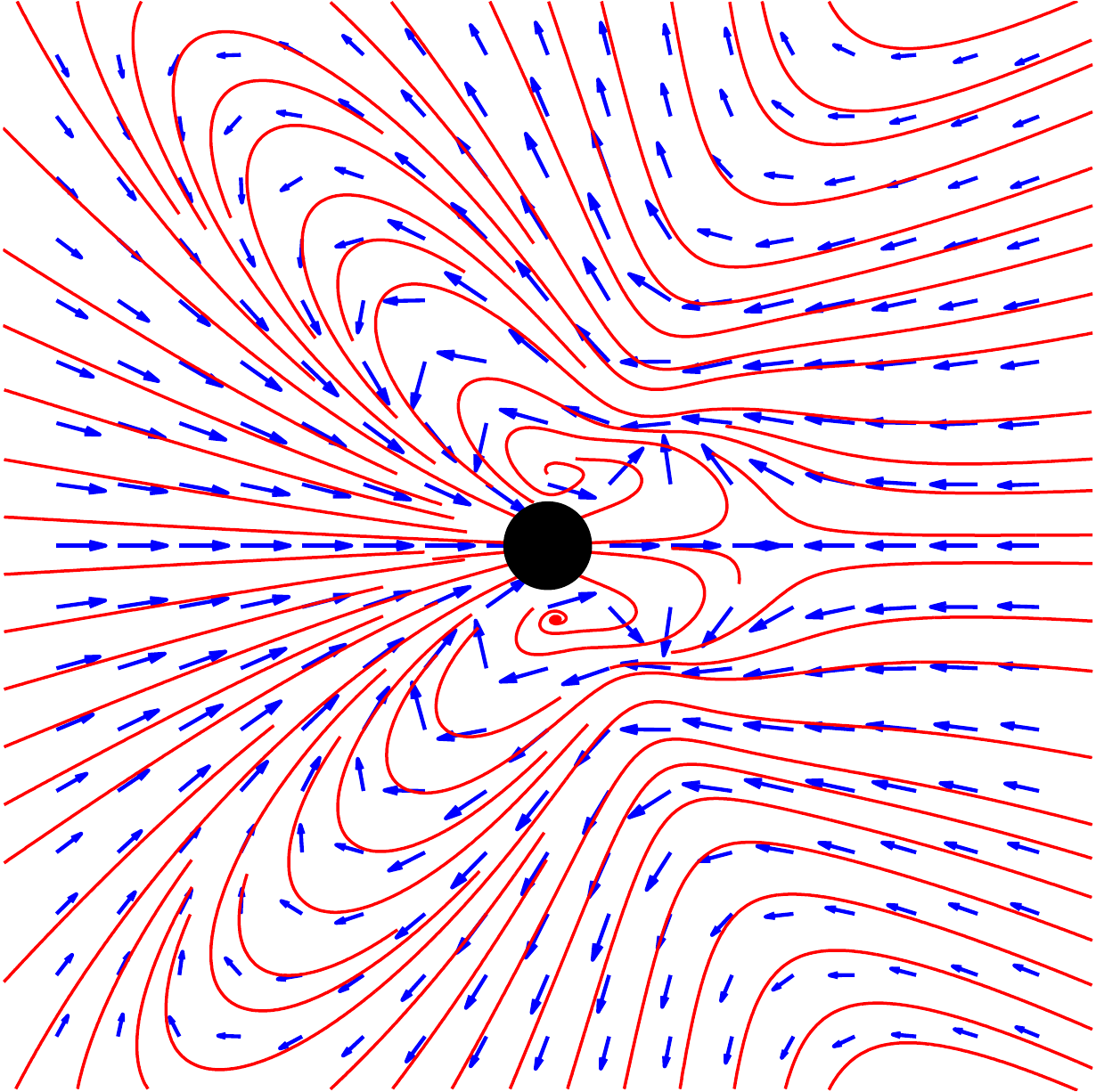}
      \caption{}
    \end{subfigure}
    \begin{subfigure}[b]{.24\textwidth}
      \includegraphics[width=\textwidth]{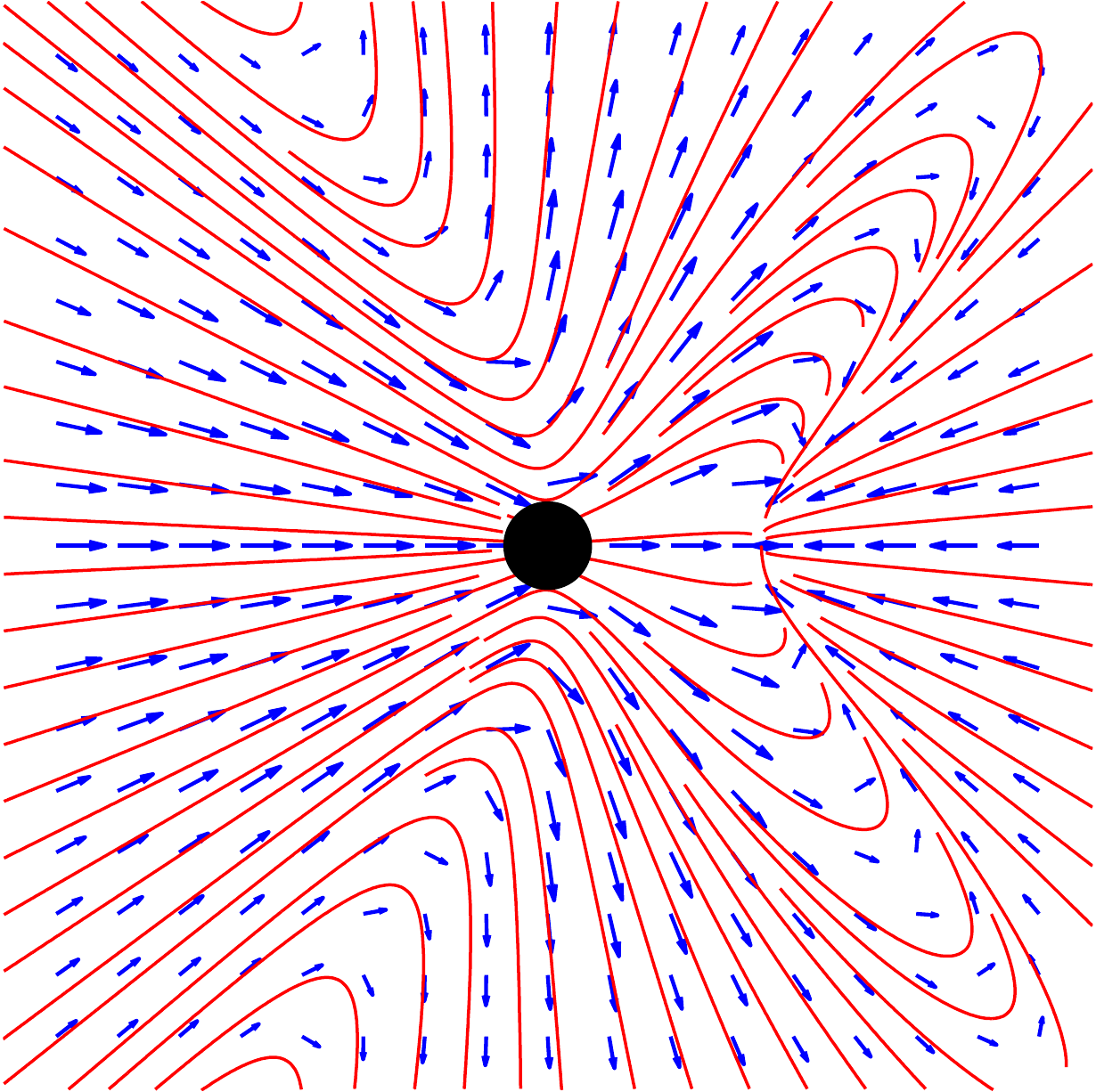}
      \caption{}
    \end{subfigure}
    \begin{subfigure}[b]{.24\textwidth}
      \includegraphics[width=\textwidth]{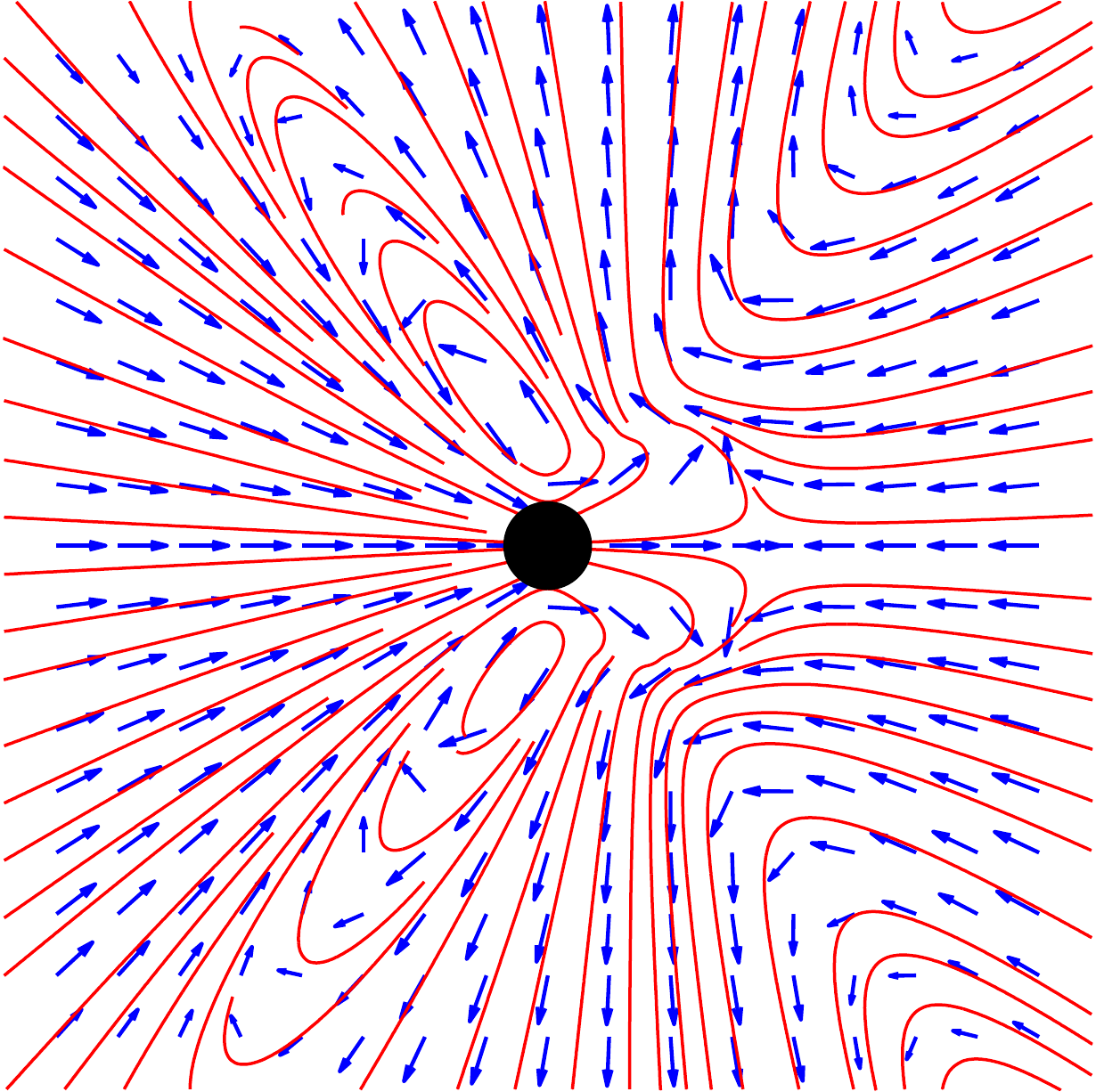}
      \caption{}
    \end{subfigure}
    \begin{subfigure}[b]{.24\textwidth}
      \includegraphics[width=\textwidth]{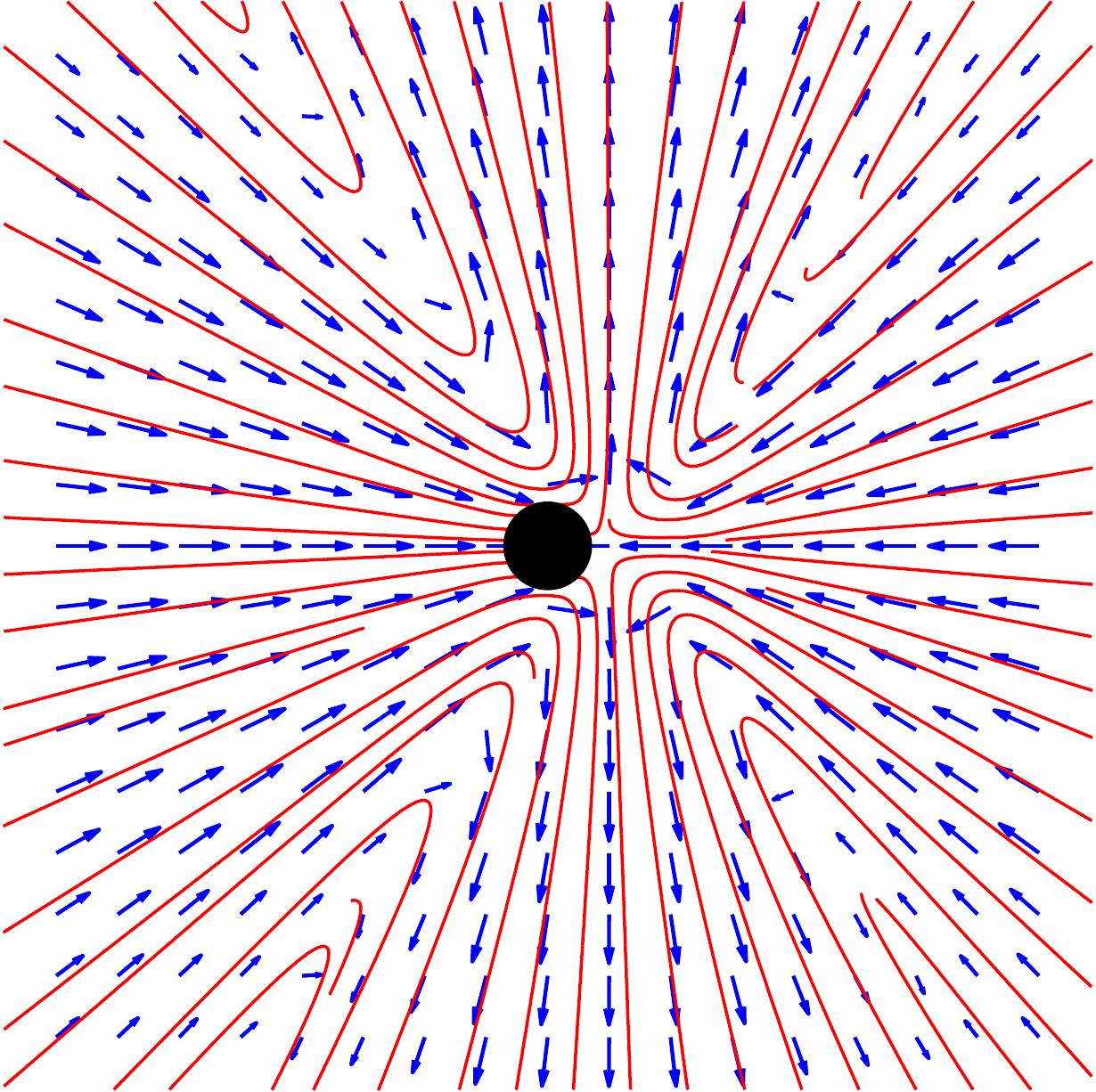}
      \caption{}
    \end{subfigure}
  \end{center}
  \caption{Time-averaged velocity fields of a 3 Stokeslet model similar to that of~\cite{Drescher2010} for: (a) the in-plane cross section ($xy$-plane containing flagella; compare to~\cite[Fig.~2(a)]{Guasto2010}), (b) out of plane cross section ($xz$-plane), and (c) the azimuthally averaged flow. (d) Time-averaged velocity field of the dumbbell model.}
  \label{fig:streamlineComparison}
\end{figure}

Axially-symmetric models involving two entities, such as this one, are often referred to as dumbbell models. Figure~\ref{fig:streamlineComparison} compares the streamlines of our axially-symmetric model to a 3-Stokeslet model that matches experimental results~\cite{Drescher2010,Guasto2010} for \emph{C.\ reinhardtii}, which lacks this symmetry. From the point of view of drift, our main interest is the interplay of a solid no-slip surface and moving flagella, so it is crucial to get the surface right, but less important to represent the flagella accurately. The actual flagella separate and partially wrap around the swimmer, which cannot occur in a dumbbell model; but we do match the swimmer's size, velocity, and oscillation frequency.

The swimmer moves at a mean swimming speed~$\Uc$ along the~$\xc$-axis. In the comoving frame the body sphere is located at~$(\AA(\time),0,0)$, with fixed radius~$\RR$ (Fig.~\ref{fig:swimmer}). The flagellar Stokeslet is located at~$(\aa(\time),0,0)$, with effective radius~$\rr(\time)$.  By axial symmetry, the drag on the sphere has the form~$\Fv=(\F,0,0)$. Fax\'en's Law for the drag on the body sphere~\cite{GuazzelliMorris} gives
\begin{equation}
  \F=6\pi\mu\RR\l(1+\tfrac16\RR^2\nabla^2\r)
  \l.\uc_{\mathrm{flag}}(\rv)\r|_{\rv=(\AA,0,0)}
  -6\pi\mu\RR\,\big(\Uc+\dA\big),
  \label{eq:Faxenlaw}
\end{equation}
where~$\mu$ is the dynamic viscosity of the fluid and~$\uc_{\mathrm{flag}}(\rv)=\uv_{\mathrm{flag}}(\rv)\cdot\xuv$ is the~$\xc$-component of the velocity due to the flagellar Stokeslet (see Section~\ref{subsec:regStokeslet}). Table~\ref{tab:notation} lists the variables and their meaning.

\begin{table}
  \caption{Notation used in the paper.}
\label{tab:notation}
\begin{center}
\begin{tabular}{cl}
\hline\hline
notation & description \\[2pt]
\hline
$\RR$ & swimmer body radius \\
$\rr(\time)$ & effective flagellar Stokeslet radius \\
$(\aa(\time),0,0)$ & position of flagellar Stokeslet in comoving frame \\
$(\AA(\time),0,0)$ & position of swimmer's body in comoving frame \\
$\Omega$ & flagellar angular frequency \\
$\tau=2\pi/\Omega$ & period of flagellar cycle \\
$(\Uc,0,0)$ & swimmer mean velocity \\
$\f(\time)$ & force on fluid due to flagellar Stokeslet \\
$\F(\time)$ & force on swimmer's body due to flow \\
$\beta(\time)$ & stresslet coefficient; see Section~\ref{sec:farfield} \\
$\lambda(\time)$ & swimming path length ($=\Uc\time$) \\
\hline\hline
\end{tabular}
\end{center}
\end{table}

A neutrally buoyant swimmer in the Stokes regime leads to no net force on the fluid; hence, $\F=\f$, where~$\F$ is the force on the sphere and
\begin{equation}
  \f=6\pi\mu\rr\,(\Uc+\da)
  \label{eq:Stokesletforce}
\end{equation}
is the force due to the flagellar Stokeslet. Combining~\eqref{eq:Faxenlaw} and~\eqref{eq:Stokesletforce} gives us a differential equation relating the position of the swimmer's body~$\AA(\time)$ and of the flagellar Stokeslet~$\aa(\time)$:
\begin{equation}
  \dA=-\Uc
  +\l[\l(1+\tfrac16\RR^2\nabla^2\r)
  \uc_{\mathrm{flag}}(\rv)\r]_{\rv=(\AA,0,0)}
  -(\Uc+\da)\,\rr/\RR.
  \label{eq:ODEforA}
\end{equation}

In order to solve~\eqref{eq:ODEforA} for~$\AA(\time)$, given $\aa(\time)$, we
must impose some additional constraints.  In a comoving frame traveling at the
mean swimming speed~$\Uc$, the time-averaged velocities of the spheres must
vanish:
\begin{equation}
  \tavg{\dA} = \tavg{\da} = 0,
  \label{eq:tavgAa}
\end{equation}
where~$\tavg{\cdot}$ denotes the average over a time period~$\tau=2\pi/\Omega$.  The simplest time-dependence we can put on the flagellar Stokeslet is
\begin{equation}
  \aa(\time) = \AA(0) + \aa_0 + \aa_1\cos\Omega\time,\qquad
  \rr(\time) = \rr_0 + \rr_1\sin\Omega\time.
  \label{eq:dynamics}
\end{equation}
These are out of phase to mimic the swimmer's power and recovery strokes.  Note that~$\tavg{\da} = 0$ since it is assumed periodic.  We've also defined~$\aa(\time)$ relative to~$\AA(0)$, since Eq.~\eqref{eq:ODEforA} is invariant under a shift of~$\aa(\time)$ and~$\AA(\time)$ by the same constant.

\begin{figure}
  \begin{center}
    \includegraphics[width=.5\textwidth]{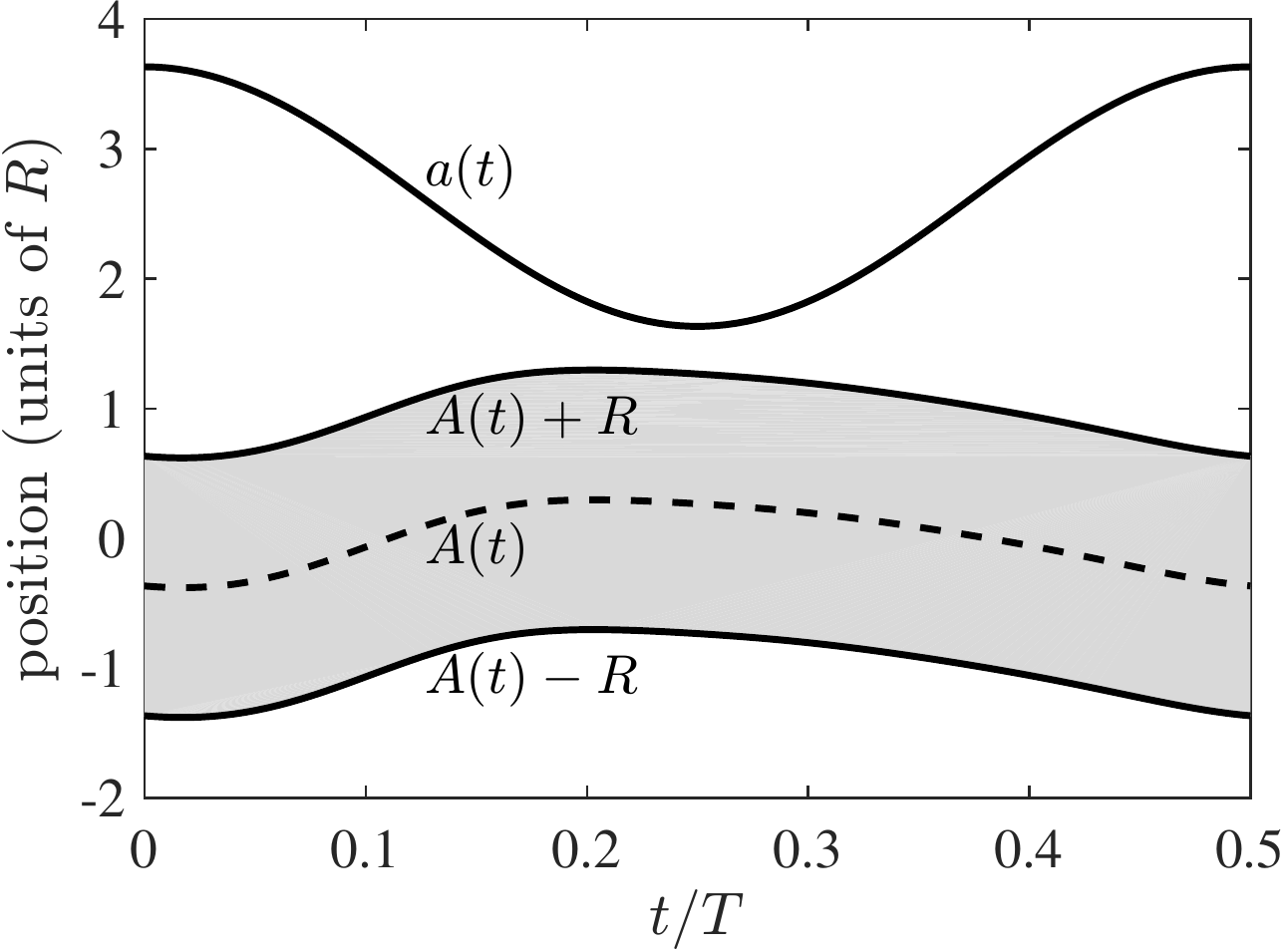}
    \caption{The position of the body sphere center~$\AA(\time)$ and the flagellar Stokeslet~$\aa(\time)$, also showing the extent of the body.   These are in the comoving frame during one full period ($\tau=0.5\timescale$), plotted along the non-dimensionalized axes using the scales in Table~\ref{tab:pset}.}
    \label{fig:bodyandStokesletposition}
  \end{center}
\end{figure}

A few observations on the strategy for solving for~$\AA(\time)$ are in order.  Equation~\eqref{eq:ODEforA} with the constraints~\eqref{eq:tavgAa} form a nonlinear eigenvalue problem for~$\AA(\time)$ and the mean swimming speed~$\Uc$ (the eigenvalue).  Only in very special cases will an analytic solution be available, so we proceed numerically.  We use a shooting method: we start with a guess for~$\Uc$, then integrate~\eqref{eq:ODEforA} until time~$\tau$, with initial condition~$\AA(0)=0$.  We then iterate by varying~$\Uc$ until~$\AA(0)=\AA(\tau)$ (using Matlab's \texttt{fzero}).  The choice~$\AA(0)=0$ is arbitrary, and it proves more convenient to subtract the average of~$\AA$ from~$\aa(\time)$ and~$\AA(\time)$ to make~$\tavg{\AA}=0$. See Fig.~\ref{fig:bodyandStokesletposition} for a plot of~$\AA(\time)$ and~$\aa(\time)$ over one full period in the comoving frame, using the physical parameters listed in Table~\ref{tab:pset} and described below.

We select the parameters of our model according to~\cite{Guasto2010,Friedrich2012,Leptos2009} (and references therein). We take an effective spherical body radius of $\RR=4\,\microm$, and the number of flagellar beats/strokes per second, $f_b=50\,\Hertz$ (or~$\Omega=100\pi\,\radian/\second$). The flagella are represented by a single Stokeslet located at~$\aa(\time)$ with effective radius~$\rr(\time)$, with time-dependence as in Eq.~\eqref{eq:dynamics}. We pick the free variables~$\aa_0$, $\aa_1$, $\rr_0$, $\rr_1$ in order to yield a mean swimming velocity close to~$100\,\microm/\second$, while also trying to match the oscillating drag due to the beating flagella.  We introduce a length scale $L=4\,\microm$ and a time scale $T=1/25\,\second$ to non-dimensionalize our system, yielding a swimmer with unit body radius and with a stroke period of~$1/2$.  These parameters and and their non-dimensionalized values are collected in Table~\ref{tab:pset}.

\begin{table}
  \caption{Physical parameters chosen to be of the same order as for \emph{C.\ reinhardtii}.  The non-dimensionalization uses a length scale~$\lengthscale=4\,\microm$ and time scale~$\timescale=1/25\,\second$.}
\label{tab:pset}
\begin{center}
\begin{tabular}{cccl}
\hline\hline
notation & value & dimensionless & description \\[2pt]
\hline
$\Uc$ & $98\,\microm/\second$ & $0.98$ &
  swimming speed \\
$\RR$ & $4\,\microm$ & $1$ & body radius \\
$f_b$ & $50\,\Hertz$ & $2$ & flagellar beat frequency \\
$\tau$ & $1/f_b$ & $1/2$ & period of flagellar cycle \\
$\Omega$ & $2\pi f_b$ & $4\pi$ & flagellar angular frequency \\
$\rr_0$ & $3.2\,\microm$ & $0.8$ & average effective flagellar radius \\
$\rr_1$ & $2\,\microm$ & $0.5$ & flagellum's radial oscillation \\
$\aa_0$ & $12\,\microm$ & $3$ & relative position of flagellar Stokeslet \\
$\aa_1$ & $4\,\microm$ & $1$ & oscillation amplitude of flagellar position \\
\hline\hline
\end{tabular}
\end{center}
\end{table}


\subsection{Flow field}
\label{subsec:flowField}

The velocity field due to a translating sphere involves a Stokeslet and a source doublet:
\begin{equation}
  \uv_{\mathrm{sphere}}(\rv)=
  6\pi\mu\RR\,\big(\Uc+\dA\big)\,\xuv\cdot\l(
  1 + \tfrac16\RR^2\lapl\r)\Os(\rv^*),
  \label{eq:translatingSphere}
\end{equation}
where~$\rv^*=\rv-\AA\xuv$ and~$\Os(\rv)$ is the Oseen tensor
\begin{equation}
  \Os(\rv)
  =\frac{1}{8\pi\mu\lVert\rv\rVert}\l(\id+\frac{\rv\rv}{\lVert\rv\rVert^2}\r).
  \label{eq:Oseen}
\end{equation}
The velocity due to the flagellar Stokeslet is
\begin{equation}
  \uv_{\mathrm{flag}}(\rv) = \f\xuv\cdot\Os(\rv-\aa\,\xuv).
  \label{eq:Stokesletflowfield}
\end{equation}
If we add the flagellar Stokeslet, we need to include images inside the sphere to preserve the no-slip boundary condition, as described by Oseen~\cite{Oseen1927}.  Here we use the simplified form for a sphere and Stokeslet that are axisymmetrically aligned along the direction of motion~\cite{Maul1996,Fuentes1988}:
\begin{multline}
  \uv_{\mathrm{image}}(\rv)=
  -\tfrac12\l(3\RRAsep-\RRAsep^3\r)\f\xuv\cdot\Os(\rv-\StImagev)
  +\RR\l(\RRAsep^2-\RRAsep^4\r)(\f\xuv\xuv\cdot\nabla)\cdot\Os(\rv-\StImagev) \\
  -\tfrac14\RR^2\RRAsep\l(1-\RRAsep^2\r)^2\f\xuv\cdot\nabla^2\Os(\rv-\StImagev),
  \label{eq:imageSystem}
\end{multline}
where~$\RRAsep=\RR/\Asep$,~$\Asep=\aa-\AA$ is the separation between the flagellar Stokeslet and the center of the swimmer's body, and~$\StImagev=\l(\RR^2/\Asep+\AA\r)\xuv$ is the location of the image singularities in the comoving frame.

Lastly we add an ambient flow in the comoving frame to get the full velocity field of our model:
\begin{equation}
  \uv_{\mathrm{comov}}(\rv) =
  -\Uc\xuv+\uv_{\mathrm{flag}}(\rv)+\uv_{\mathrm{sphere}}(\rv)
  +\uv_{\mathrm{image}}(\rv).
  \label{eq:fullFlowField}
\end{equation}
This is related to the velocity field in the lab (fixed) frame by
\begin{equation}
  \uv_{\mathrm{lab}}(\rv) = \uv_{\mathrm{comov}}(\rv-\Uc\time\,\xuv)+\Uc\xuv.
  \label{eq:labFrame}
\end{equation}
Recall from Section~\ref{subsec:model} that many of the parameters, such as those defined in~\eqref{eq:dynamics}, will have a time-dependence that we did not indicate explicitly in the velocity fields above.


\subsection{Regularization of the flagellar Stokeslet}
\label{subsec:regStokeslet}

One of the main motivations for our dumbbell model is to account for the rigid no-slip surface of the body, since it can lead to stickiness of fluid particles~\cite{Thiffeault2010b}.  To simplify the model as much as possible, we used a point-singularity representation for the flagellum.  Since we want to simulate the advection of particles by our swimmer, it is wise to regularize the flagellum in order to avoid infinite velocities inside the fluid.  Here we pick the regularization from the analytic model in~\citet{HernandezOrtiz2007}. The flow field of the regularized flagellar Stokeslet in the comoving frame is
\begin{equation}
  \uv_{\mathrm{flag}}(\rv)
  = \f\xuv\cdot\Os^{\xi}(\rv-\aa\,\xuv).
  \label{eq:regStokesletflowfield}
\end{equation}
Here~$\Os^{\xi}(\rv)$ is a regularized Oseen tensor,
\begin{equation}
  \Os^\xi(\rv)
  = \Os(\rv)\erf(\xi\lVert\rv\rVert)
  +\frac{1}{8\pi\mu}\l(\id-\frac{\rv\rv}{\lVert\rv\rVert^2}\r)
  \frac{2\xi}{\sqrt{\pi}}\,\ee^{-\xi^2\lVert\rv\rVert^2}
  \label{eq:regOseen}
\end{equation}
where~$\Os(\rv) = \Os^\infty(\rv)$ is the standard (unregularized) Oseen tensor~\eqref{eq:Oseen}.  The velocity field of a regularized Stokeslet, $\uv^{\xi}(\rv)=\fv\cdot\Os^{\xi}(\rv)$, satisfies Stokes equation
\begin{equation}
  -\grad\pres + \mu\lapl\uv^{\xi} = -\fv\delta^{\xi}(\rv),\qquad \div\uv^{\xi} = 0,
\end{equation}
where
\begin{equation}
  \delta^\xi(\rv)=\frac{\xi^3}{\pi^{3/2}}\l(\tfrac52-\xi^2\lVert\rv\rVert^2\r)
  \ee^{-\xi^2\lVert\rv\rVert^2}
\end{equation}
is a suitably-chosen regularized delta function~\cite{HernandezOrtiz2007}.

The variable~$\xi$ is a regularization parameter, with units of inverse length. In the limit~$\xi\to\infty$, we recover the unregularized Stokeslet.  We choose the regularization scale~$\xi^{-1}=\tfrac14\rr_0$, a value smaller than the minimum effective flagellar radius,~$\rr(\time)$.   This is small enough to ensure that the solution for~$\AA(\time)$ is essentially unaffected by the regularization.


\section{Numerical integration for a single swimmer}
\label{sec:numerical}

As the swimmer moves, it displaces fluid particles.  The net nonzero displacement of fluid particles after the swimmer has passed is often referred to as \emph{Darwin drift} \cite{Maxwell1869,Darwin1953,Lighthill1956}, to distinguish it from Stokes drift due to wave motion.  Here we will use the more precise word `displacement' when referring to particle drift.  The particle displacements are obtained by computing the fluid particle trajectories, and in this section we do so using numerical integration.  In Sections~\ref{sec:nearfield}--~\ref{sec:farfield} we will derive features of the particle displacements using asymptotic analysis.

We assume idealized fluid particles whose position~$\rvp$ obeys
\begin{equation}
  \dot\rvp=\uv(\rvp,\time),
  \qquad
  \rv_0 = (\xc_0,\yc_0,0),
  \label{eq:dotrvp}
\end{equation}
where we set~$\zc_0=0$ without loss of generality by exploiting the axial symmetry. In the lab (fixed) frame we use Eq.~\eqref{eq:labFrame} on the right-hand side of Eq.~\eqref{eq:dotrvp} and include the time-dependence of all the parameters.  In the following sections we examine particle \emph{paths}, as given by~$\rvp(\time)$, and particle \emph{displacements},
\begin{equation}
  \Delta_\lambda(\xc_0,\yc_0) = \lVert\rvp - \rv_0\rVert,
  \qquad
  \lambda = \Uc\time,
\end{equation}
where~$\lambda$ is the swimmer's path length. It is well-known that particles can have paths that undergo large excursions, and yet have relatively small displacements~\cite{Maxwell1869,Darwin1953}. At moderate and far distances from the swimmer, this near-closure is generic for potential and viscous flows~\cite{Dunkel2010}.  We integrate Eq.~\eqref{eq:dotrvp} numerically with Matlab's \texttt{ode45}, using the non-dimensionalized values in Table~\ref{tab:pset}.


\subsection{Particle paths}
\label{subsec:partpath}

\begin{figure}
  \begin{center}
    \begin{subfigure}[b]{.5\textwidth}
      \includegraphics[height=.24\textheight]{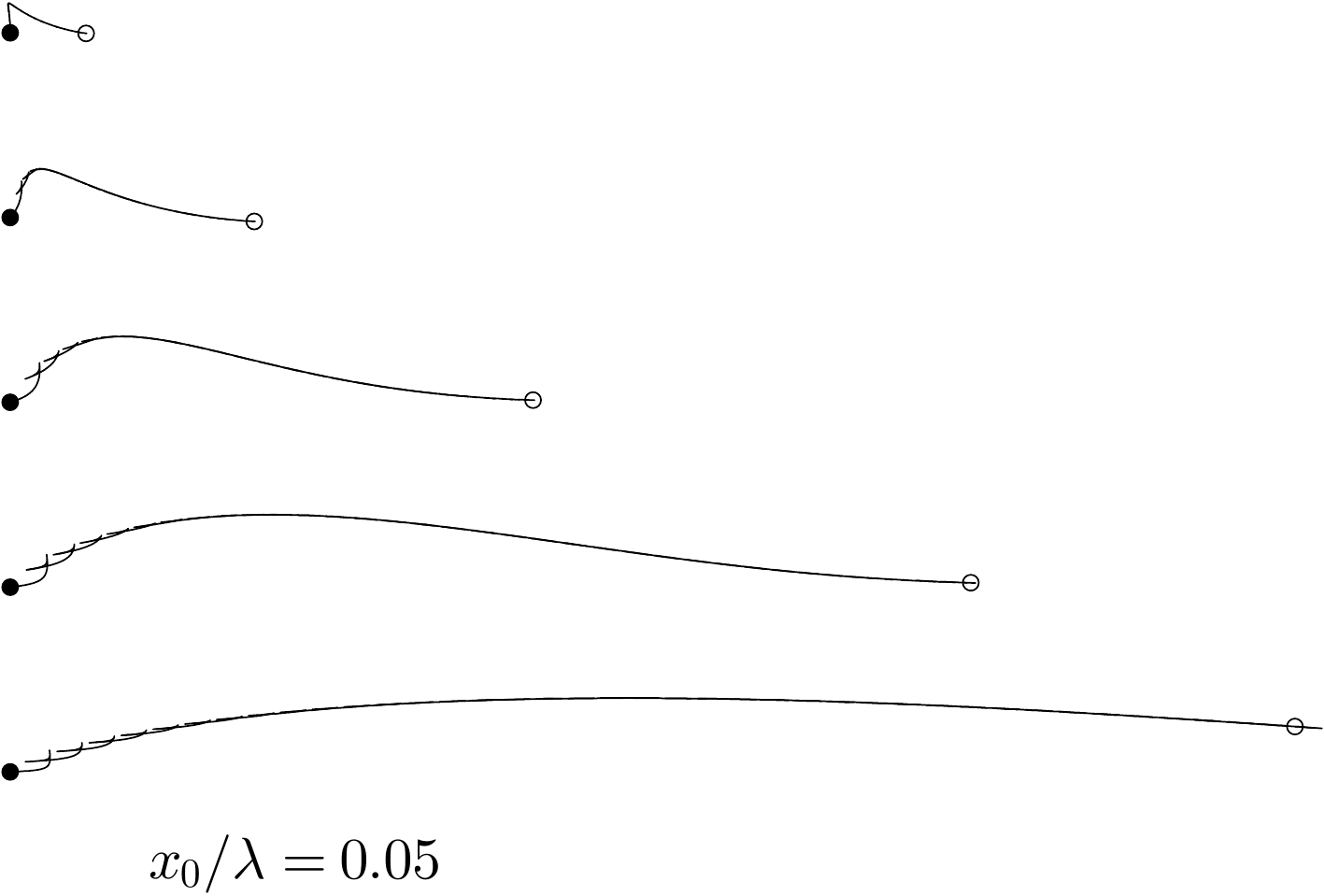}
      \caption{}
    \end{subfigure}
    \begin{subfigure}[b]{.26\textwidth}
      \includegraphics[height=.24\textheight]{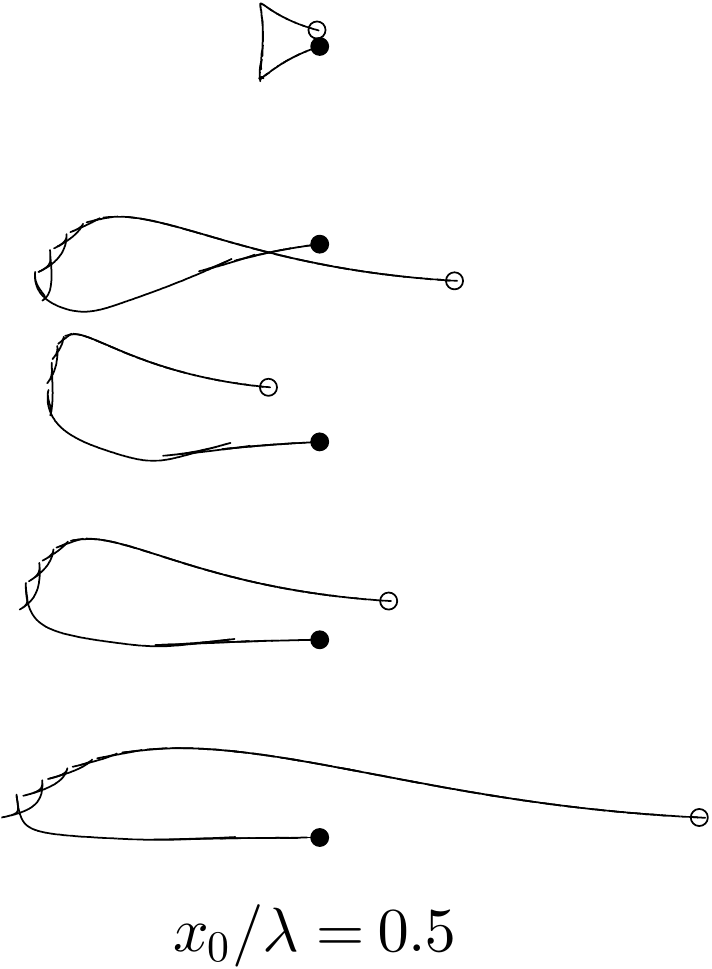}
      \caption{}
    \end{subfigure}
    \begin{subfigure}[b]{.16\textwidth}
      \includegraphics[height=.24\textheight]{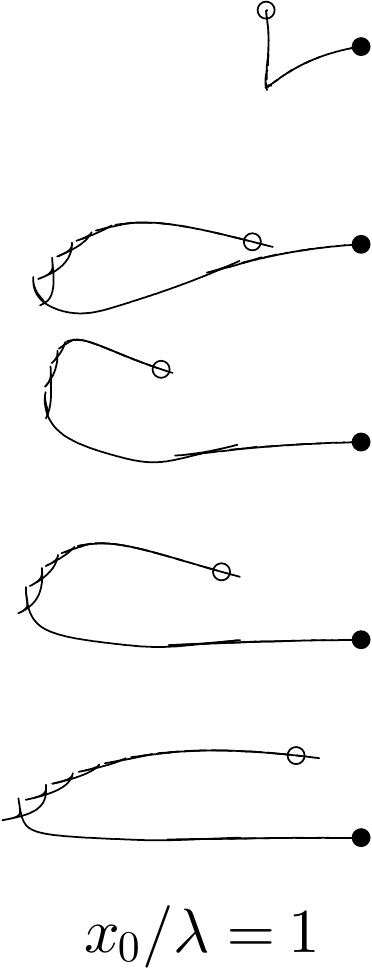}
      \caption{}
    \end{subfigure}
  \end{center}
  \caption{%
 Particle trajectories in the lab frame starting at~$(\xc_0,\yc_0,0)$.  The swimmer travels a net distance of~$\lambda=40\,\Uc\tau$. From top to bottom,~$\log(\yc_0/\lengthscale)=1,0,-1,-2,-3$. The trajectories are offset vertically for clarity. The initial position of the particles is marked by solid dots and the final position by hollow dots.  The apparent `roughness' of the paths is due to the time-dependent swimming motion.}
  \label{fig:paths_all}
\end{figure}

We first discuss the particle paths in detail, before turning to the net displacements in Section~\ref{subsec:dispint}.  We observe loop-like trajectories for distant particles as the swimmer passes by them, a result commonly found with steady swimmers~\cite{Maxwell1869,Darwin1953,Dunkel2010,Thiffeault2010b,Lin2011,Pushkin2013}. The loop-like trajectories cause the net displacement to be much smaller than the distance traveled by the particle, as seen in Fig.~\ref{fig:paths_all}(b).
The loop-like behavior is broken for particles close to the start or end of the swimmer's path (Fig.~\ref{fig:paths_all}(a,c)), as pointed out in~\cite{Lin2011}. The ends of the path are associated with sudden turns, as exhibited by \emph{E.~coli}'s run-and-tumble dynamics, but can also be related to curved trajectories~\cite{Pushkin2013b} or bounded domains~\cite{Pushkin2014}.  We will revisit particle paths far away from the swimmer in Section~\ref{sec:farfield}.


\subsection{Particle displacements}
\label{subsec:dispint}

\begin{figure}
  \begin{center}
    \includegraphics[width=\textwidth]{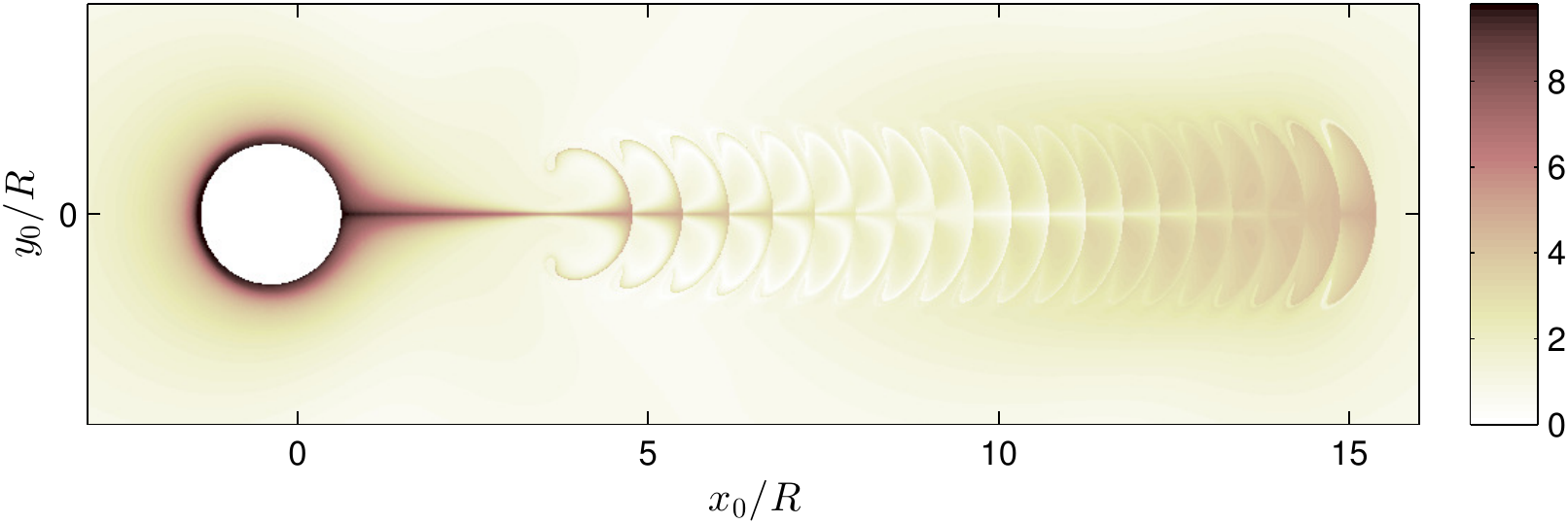}
    \caption{Plot of particle displacements,~$\Delta_{\lambda}(\xc_0,\yc_0)/\lengthscale$, as a function of initial particle position~$(\xc_0,\yc_0,0)$ for a swimmer starting at~$(\AA(0),0,0)$ and swimming for 20 periods (or a net distance of~$\lambda\approx 9.8\lengthscale$). The white disk is the initial position of the swimmer's body.}
    \label{fig:dispint01}
  \end{center}
\end{figure}

An immediate next step is to study the net displacement of each particle path. We begin by integrating an initial mesh of particles (again we assume idealized particles that follow the fluid flow). After integrating the particles, we can then calculate and plot their net displacement (Fig.~\ref{fig:dispint01}). We recover the open trajectories (and thus larger displacements) of particles located near the start and end of the finite swimming path as mentioned in the previous section.  We also see large displacements for particles that are in the path of the no-slip surface of the swimmer's body; this can be seen by the streak of large displacements along the swimming axis, immediately ahead of the sphere.

A sequence of faint stripes can also be seen in the right half of Fig.~\ref{fig:dispint01}. They first appear near the initial location of the flagellar Stokeslet and repeat almost periodically.  The spacing between stripes is roughly equal to $\Uc\tau$, the distance traveled by the swimmer in one period.  The leftmost stripe occurs at the first maximum excursion of the flagellar Stokeslet from the the swimmer's body (see Fig.~\ref{fig:bodyandStokesletposition}).

\begin{figure}
  \begin{center}
    \includegraphics[width=.7\textwidth]{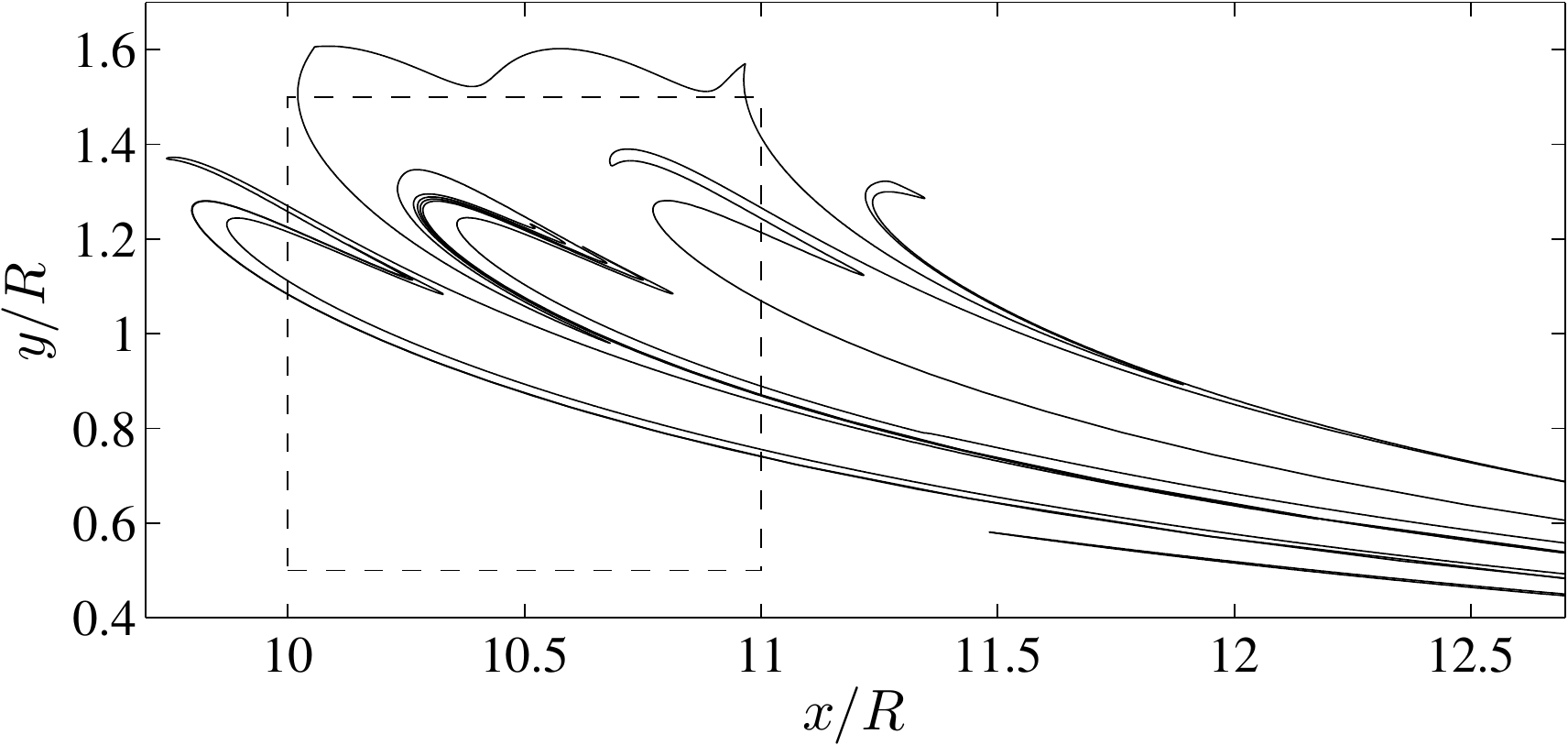}
    \caption{The swimmer starts centered at the origin and swims for 100 periods (a distance of about~$49\lengthscale$), passing through the an initial square  of fluid particles (dashed) and deforming it (solid).  The time dependence creates characteristic lobe structures.}
    \label{fig:cloud_drift}
  \end{center}
\end{figure}

It is also instructive to examine how material lines of fluid particles are displaced by the swimmer.  In Fig.~\ref{fig:cloud_drift} we take an initial square of fluid particles (dashed), located ahead of the swimmer.  The solid lines then show the eventual fate of that square as its constituent particles are displaced by the swimmer.  Notice the large amount of stretching and folding that creates `lobes,' typically associated with mixing~\cite{Aref1984,RomKedar1990,Ottino1990,Wiggins}.  Here we have a transient process, which is more appropriately analyzed using methods from transient chaos in open flows~\cite{Tel1996,Pentek1999}.  We do not carry out such an analysis here; instead we will discuss mixing in terms of the statistics of particle displacements (Section~\ref{sec:stats}).

\section{Near-field asymptotics}
\label{sec:nearfield}

The trajectories with the largest displacements commonly occur near the swimmer.  In particular, particles directly in the path of the swimmer (small~$\yc_0$) are displaced the most.  In an inviscid fluid~\cite{Thiffeault2010b} or for `squirmers'~\cite{Lin2011}, the largest displacements typically scale as~$\log\yc_0$, since they arise from particles that remain in the vicinity of stagnation points at the leading and trailing edges of the body.  For no-slip spheres, the largest displacements scale as~$1/\yc_0$~\cite{Eames2003b}, this time due to particles that remain near the no-slip rigid surface.  For our time-dependent swimmer, the situation is more complicated, since particles near the no-slip body of the swimmer are still affected by the time-dependent flagellar Stokeslet. We now model these particles in order to identify the cause of the largest particle displacements.  We find that the largest displacements still scale as~$1/\yc_0$, due to the no-slip body, but with a smaller proportionality constant than an isolated sphere because the flagellar Stokeslet pushes particles along the body.  That `constant' also depends periodically on the initial horizontal distance~$\xc_0$.

\subsection{Flow near the swimmer's body}

In a frame moving with the swimmer's body, the velocity field is very small near the no-slip surface.  A particle near that surface in the upper-half~$\xc$--$\yc$ plane has a coordinate vector of the form~$\rv = (\AA(\time) + (\RR + \delta\rc)\cos\theta)\xuv + (\RR + \delta\rc)\sin\theta\,\yuv$, where~$\delta\rc$ is small and~$0 \le \theta \le \pi$.  The `leading edge' has~$\theta=0$, and the `trailing edge' has~$\theta=\pi$.  We Taylor expand for small~$\delta\rc$ and find the tangential velocity
\begin{equation}
  \uc_\theta(\delta\rc,\theta,\time) =
  \tfrac32\frac{\delta\rc}{\RR}\sin\theta\l\{
  (\Uc + \dA)
  - \tfrac32\frac{\rr\,(\RR^2 - \Asep^2)^2(\Uc + \da)}
  {(\RR^2 + \Asep^2 - 2\RR\Asep\cos\theta)^{5/2}}
  \r\} + \Order{(\delta\rc)^2}
  \label{eq:uthetab}
\end{equation}
where~$\Asep(\time) = \aa(\time) - \AA(\time)$.  The term proportional to~$\Uc + \dA$ is the same as for a no-slip sphere in a
flow with that speed.  Using the force balance condition
Eq.~\eqref{eq:ODEforA} to eliminate~$\da$ in~\eqref{eq:uthetab}, we find after
some work
\begin{equation}
  \uc_\theta(\delta\rc,\theta,\time) =
  \tfrac32(\Uc + \dA)\frac{\delta\rc}{\RR}\sin\theta\l\{
  1 + \W(\Asep,\theta)
  \r\}.
  \label{eq:uthetab2}
\end{equation}
where we dropped terms of order~$(\delta\rc)^2$ and defined
\begin{equation}
  \W(\Asep,\theta) \ldef
  \frac{3\RR(\RR+\Asep)^2\Asep^3}{(\RR + 2\Asep)
  (\RR^2 + \Asep^2 - 2\RR\Asep\cos\theta)^{5/2}}\,.
\end{equation}
At leading order, the corresponding radial velocity component is second-order
in~$\delta\rc$:
\begin{equation}
  \uc_\rc(\delta\rc,\theta,\time) \approx
  -\tfrac32\,(\Uc + \dA)\,\frac{(\delta\rc)^2}{\RR^2}\,\cos\theta
  \l\{1 + \W(\Asep,\theta)
  + \tfrac12\tan\theta\,\pd_\theta\W(\Asep,\theta)\r\}.
  \label{eq:urb2}
\end{equation}

Given the velocity components~\eqref{eq:uthetab2} and~\eqref{eq:urb2}, is it
possible for the flow near the boundary to exhibit a `bubble' or recirculation
region, that is, a separating streamline (in a frame oscillating with the
body) other than at~$\theta=0$ or~$\pi$?  No, since this would require the two
terms in the braces in Eq.~\eqref{eq:uthetab2} to cancel for
some~$\theta=\theta_{\text{sep}}$, but~$\W(\Asep,\theta) > 0$ since $\Asep =
\aa - \AA > \RR > 0$ to avoid collision between the flagellar Stokeslet and
body.  Hence, there is no such `bubble.'  It is notable that the nonexistence
of the recirculation region is tied to the force-free condition.  The lack of
a recirculation region means that a particle initially very close to the~$\xc$
axis (small~$\yc_0$) in the swimmer's path will crawl along the entire length
of the swimmer's body.

\subsection{Two-time expansion}

The polar coordinates of a fluid particle near the swimmer's body satisfy
\begin{equation}
  \skew{10}\dot\delta\rc = \uc_\rc(\delta\rc,\theta,\time),
  \qquad
  \dot\theta = \uc_\theta(\delta\rc,\theta,\time)/\RR,
  \label{eq:drdtheta}
\end{equation}
where~$\uc_\theta$ is given by Eq.~\eqref{eq:uthetab2} and~$\uc_\rc$ by
Eq.~\eqref{eq:urb2}.  Because both~$\uc_\rc$ and~$\uc_\theta$ vanish
at~$\delta\rc=0$, a particle near the boundary moves very little at each
period~$\tau$ with respect to the swimmer's body.  This slow
motion is captured by a slow time~$\Time$ and the expansions
\begin{equation}
  \pd_\time \rightarrow \pd_\time + \eps\,\pd_\Time,\quad
  \theta = \theta_0 + \eps\,\theta_1 + \ldots,\quad
  \delta\rc = \eps\,(\delta\rc_1 + \eps\,\delta\rc_2 + \ldots),
  \label{eq:exps}
\end{equation}
where~$\eps$ is a small parameter proportional to how close the particle is to
the body.  All the quantities now a priori depend on the two times $\time$
and~$\Time$.  We now insert the expansions~\eqref{eq:exps} into~\eqref{eq:drdtheta}, use the
leading-order dependence of~$\uc_\rc$ and~$\uc_\theta$ with~$\delta\rc$, and
equate powers of~$\eps$.  At leading order in~$\eps$ this gives
\begin{equation}
  \pd_\time\delta\rc_1 = 0,
  \qquad
  \pd_\time\theta_0 = 0,
\end{equation}
so that~$\delta\rc_1(\time,\Time)=\delta\rc_1(\Time)$ and~$\theta_0(\time,\Time)=\theta_0(\Time)$.  At the next order, we obtain
\begin{equation}
  \pd_\Time\delta\rc_1 +
  \pd_\time\delta\rc_2 = \uc_\rc(\delta\rc_1,\theta_0,\time),
  \qquad
  \pd_\Time\theta_0 +
  \pd_\time\theta_1 = \uc_\theta(\delta\rc_1,\theta_0,\time)/\RR.
  \label{eq:eps1}
\end{equation}
We average~\eqref{eq:eps1} over one period in~$\time$ and impose periodicity
of~$\delta\rc_2$ and~$\theta_1$, so
that~$\tavg{\pd_\time\delta\rc_2}=\tavg{\pd_\time\theta_1}=0$:
\begin{equation}
  \pd_\Time\delta\rc_1(\Time)
  = \tavg{\uc_\rc(\delta\rc_1(\Time),\theta_0(\Time),\cdot)},
  \qquad
  \pd_\Time\theta_0(\Time)
  = \tavg{\uc_\theta(\delta\rc_1(\Time),\theta_0(\Time),\cdot)}/\RR.
  \label{eq:periodavg}
\end{equation}
The~`$\cdot$' argument indicates that we are averaging only with respect to
the last slot, holding~$\delta\rc$ and~$\theta$ fixed
(see~\eqref{eq:tavgvel} below).  These are `slow' equations that capture a
particle's drift near the boundary, using a period-averaged velocity.  To
simplify the notation, we now drop the subscripts and use~$\time$
for~$\Time$ in~\eqref{eq:periodavg}:
\begin{equation}
  \skew{10}\dot\delta\rc
  = \tavg{\uc_\rc(\delta\rc,\theta,\cdot)},
  \qquad
  \dot\theta
  = \tavg{\uc_\theta(\delta\rc,\theta,\cdot)}/\RR,
  \label{eq:periodavg2}
\end{equation}
with
\begin{equation}
  \tavg{\uc_\rc(\delta\rc,\theta,\cdot)} =
  \frac{1}{\tau}\int_{0}^{\tau}
  \uc_\rc(\delta\rc,\theta,\stime)\dint\stime,
  \qquad
  \tavg{\uc_\theta(\delta\rc,\theta,\cdot)} =
  \frac{1}{\tau}\int_{0}^{\tau}
  \uc_\theta(\delta\rc,\theta,\stime)\dint\stime.
  \label{eq:tavgvel}
\end{equation}
The particle displacement equations~\eqref{eq:periodavg2} are time-averaged in
the sense that they now only depend on~$\time$ through the change
in~$\delta\rc(\time)$ and~$\theta(\time)$.

\begin{figure}
  \begin{center}
    \includegraphics[width=.6\textwidth]{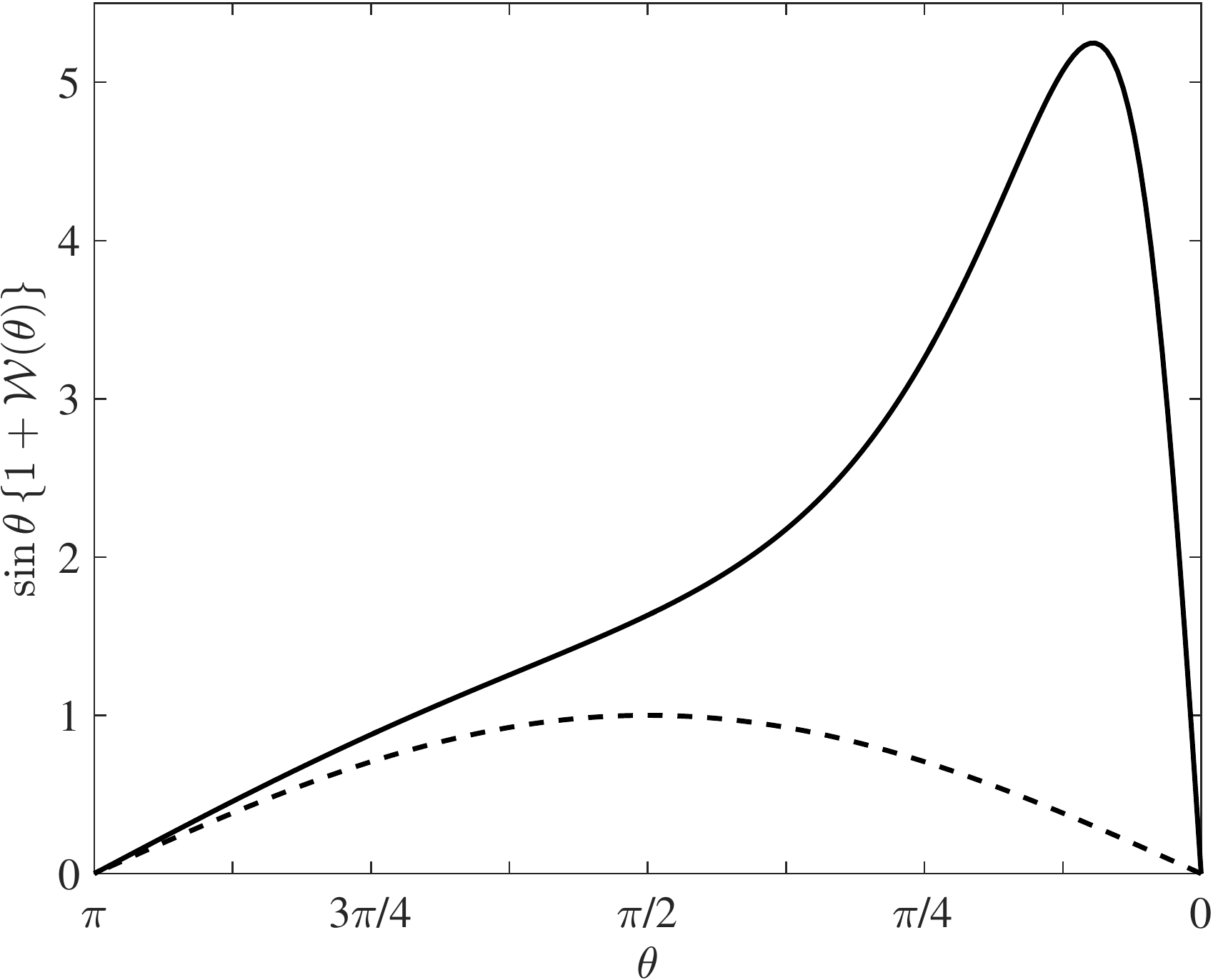}
    \caption{The averaged tangential speed \eqref{eq:Uavg}, after dropping the
      lead coefficient.  The dashed line is for an isolated rigid sphere
      ($\Wavg\equiv0$) moving at the same speed.  The front of the swimmer
      is to the right.}
    \label{fig:Wavg}
  \end{center}
\end{figure}

Let's evaluate~$\tavg{\uc_\theta}$. From~\eqref{eq:uthetab2}, we have
\begin{equation}
  \tavg{\uc_\theta(\delta\rc,\theta,\cdot)} =
  \tfrac32\frac{\Uc}{\RR}\,\delta\rc\,\sin\theta\l\{1
  +
  \Wavg(\theta)\r\}.
  \label{eq:Uavg}
\end{equation}
where
\begin{equation}
  \Wavg(\theta) \ldef
  \frac{1}{\tau}\int_0^\tau
  (1 + \dA(\stime)/\Uc)\, \W(\Asep(\stime),\theta)
  \dint\stime.
  \label{eq:Wavg}
\end{equation}
The integral~\eqref{eq:Wavg} is straightforward to evaluate numerically.  In
Fig.~\ref{fig:Wavg} we compare the averaged velocity~\eqref{eq:Uavg} (solid
line) to an isolated no-slip sphere ($\Wavg\equiv0$, dashed line) for our
reference parameter values.  The averaged velocity is much larger on the front
side of the swimmer (right) due to the effect of the flagellar Stokeslet.  As
we shall see below, this implies paradoxically that fluid particles are
displaced \emph{less} in the fixed lab frame, since their residence time in
the boundary region is shorter than for an isolated no-slip sphere (see
Section~\ref{sec:nearnetdisp}).  Put another way, the swimmer's body is less
`sticky' than an isolated no-slip sphere.  (This difference is partially
mitigated by particles coming closer to the swimmer's body than for a no-slip
sphere, see Eq.~\eqref{eq:rrtheta}.)

\subsection{Averaged streamline}

\begin{figure}
  \begin{center}
    \includegraphics[width=.75\textwidth]{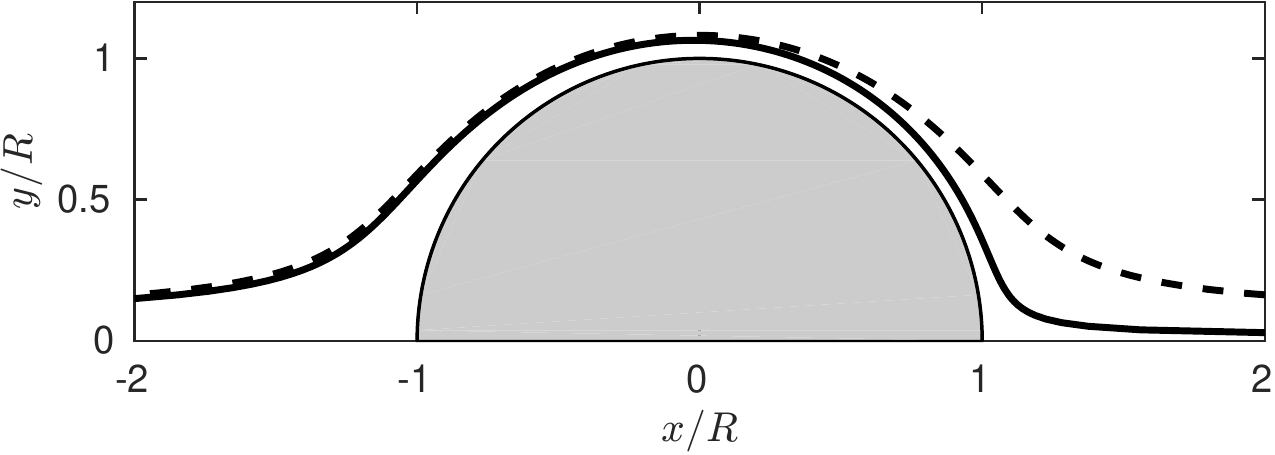}
    \caption{The averaged streamline \eqref{eq:rrtheta} (solid line) is closer
      to the swimmer's body than for an isolated no-slip sphere ($\Wavg\equiv0$,
      right) moving at the same speed.  The streamline is closest to the body
      at the front of the swimmer (right).}
    \label{fig:slavg}
  \end{center}
\end{figure}

For an axisymmetric flow, we can define a streamfunction~$\psi(\rc,\theta)$ such that
\begin{equation}
  \uc_\rc(\rc,\theta) = \frac{1}{\rc^2\sin\theta}\,\frac{\pd\psi}{\pd\theta}\,,
  \qquad
  \uc_\theta(\rc,\theta) = -\frac{1}{\rc\sin\theta}\,\frac{\pd\psi}{\pd\rc}\,.
  \label{eq:usfaxial}
\end{equation}
Using this with~$\rc=\RR+\delta\rc$ we can find a streamfunction for the
averaged flow~\eqref{eq:Uavg}:
\begin{equation}
  \psi(\RR+\delta\rc,\theta) =
  -\tfrac34\,\Uc\,(\delta\rc)^2\sin^2\theta\l\{1
  + \Wavg(\theta)\r\} + \Order{\delta\rc^3},
  \label{eq:psiavg}
\end{equation}
valid in the vicinity of the swimmer's body to leading order in~$\delta\rc$.
The streamfunction far from the swimmer is
\begin{equation}
  \psi_\infty(\rc,\theta)
  = -\tfrac12\Uc\rc^2\sin^2\theta = -\tfrac12\Uc\yc^2,
  \qquad \rc \gg \RR,
\end{equation}
which corresponds to the steady flow to the left.  The equation for the
`average streamline' where a particle ends up at~$\yc_1$ after the swimmer has
passed is then obtained by
setting~$\psi(\RR+\delta\rc,\theta) = \psi_\infty(\yc_1) =
-\tfrac12\Uc\yc_1^2$, which gives
\begin{equation}
  \tfrac32\,(\delta\rc)^2\sin^2\theta\l\{1
  + \Wavg(\theta)\r\}
  = \yc_1^2.
  \label{eq:avgslcond}
\end{equation}
We then solve this for~$\delta\rc(\theta)$:
\begin{equation}
  \delta\rc(\theta)
  = \sqrt{\tfrac23}\,\frac{\yc_1}{\sin\theta}\l\{1
  + \Wavg(\theta)\r\}^{-1/2}.
  \label{eq:rrtheta}
\end{equation}
With~$\Wavg\equiv0$ we recover the streamline for an isolated no-slip sphere
in a constant flow (see Fig.~\ref{fig:slavg}).  The term~$\Wavg(\theta)$ is
positive, so the swimmer's averaged streamline is always closer to the body
than for the equivalent isolated no-slip sphere.  The difference between a
streamline for the isolated no-slip sphere and for the swimmer is most
pronounced at~$\theta=0$, as expected since this is the side of the flagellar
Stokeslet.

\begin{figure}
  \begin{center}
    \includegraphics[width=.75\textwidth]{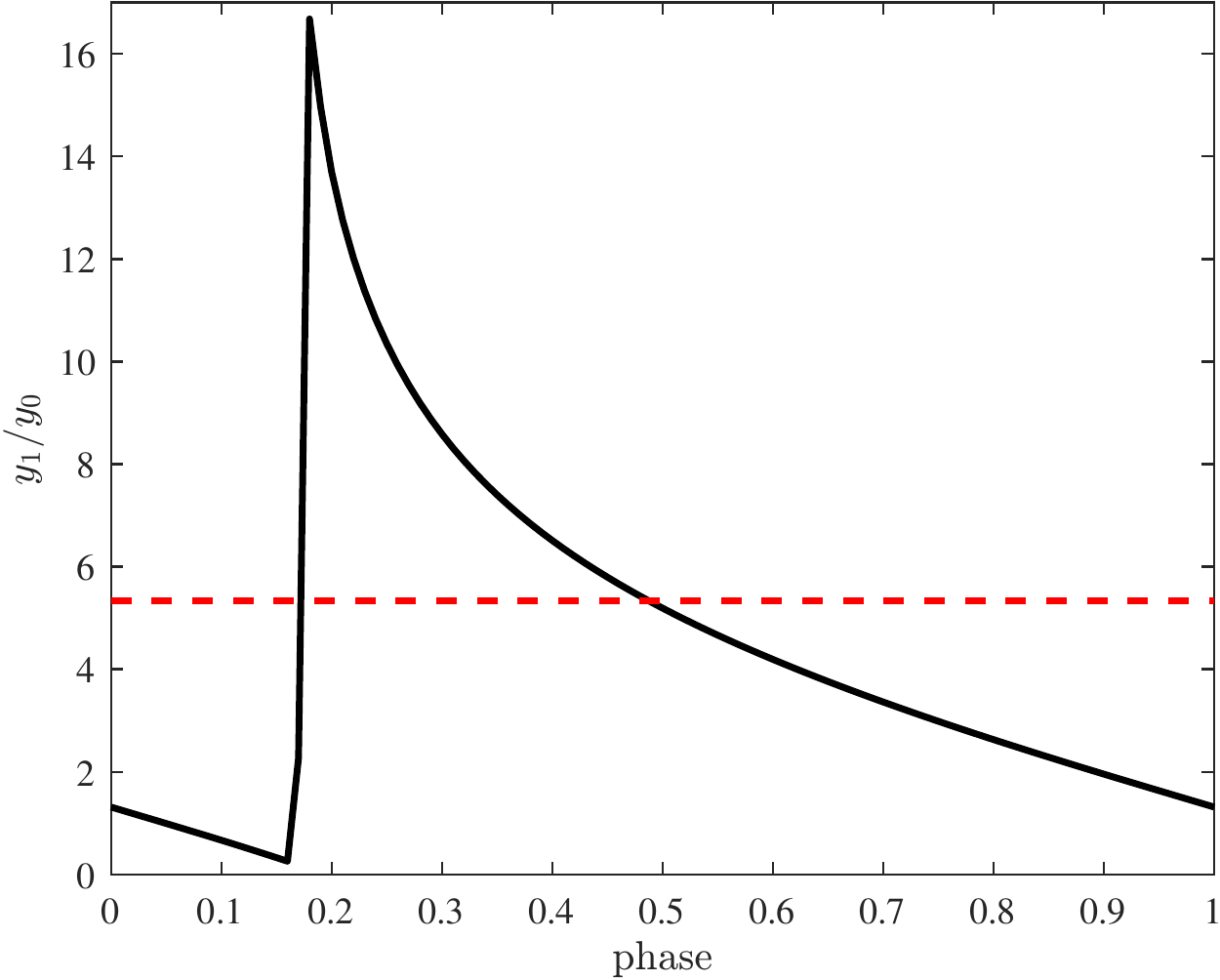}
    \caption{Ratio of final to initial~$\yc$ value, representing the jump in
      streamline due to the time dependence.  The phase is expressed as a
      fraction of the period~$\tau$.  The dashed line is the time-averaged
      expression~\eqref{eq:533}.}
    \label{fig:puller_y1Cy0}
  \end{center}
\end{figure}

In a steady flow, a particle that starts at~$\yc_0$, far ahead of the swimmer,
returns to~$\yc_0$ after the swimmer has passed.  Because of the time
dependence, the streamfunction can change value.  We can estimate this change
from~\eqref{eq:rrtheta}:
\begin{equation}
  \frac{\yc_1}{\yc_0} \approx
  \frac{[(\RR + \delta\rc(\theta))\sin\theta]_{\theta=\pi}}
  {[(\RR + \delta\rc(\theta))\sin\theta]_{\theta=0}}
  =
  \sqrt{\frac{1 + \Wavg(0)}{1 + \Wavg(\pi)}}
  \approx 5.33.
  \label{eq:533}
\end{equation}
However, this is at best a rough approximation, since it involves
taking~$\delta\rc$ to infinity when it should be small, as well as being based
on the time-averaged velocity.  Figure~\ref{fig:puller_y1Cy0} shows that the
ratio of the final to initial~$\yc$ depends on the phase of the flagellar
Stokeslet.  The value~$5.33$ from~\eqref{eq:533} (dashed line) does sit
roughly in the middle.

The jump is caused by particles coming near the regularized Stokeslet singularity. As particles first interact with the swimmer, some particles end up on one side or the other of the flagellar Stokeslet. Those who remain in front of the Stokeslet take approximately one more swimming stroke to move around the flagellar Stokeslet, thus becoming separated from neighboring particles that were on the other side (the cause of the folds in Fig.~\ref{fig:cloud_drift}). This effect can be seen by either varying the initial particle position or by changing the initial flagellar phase while keeping everything else constant.

\subsection{Net displacement}
\label{sec:nearnetdisp}

\begin{figure}
  \begin{center}
    \includegraphics[width=.75\textwidth]{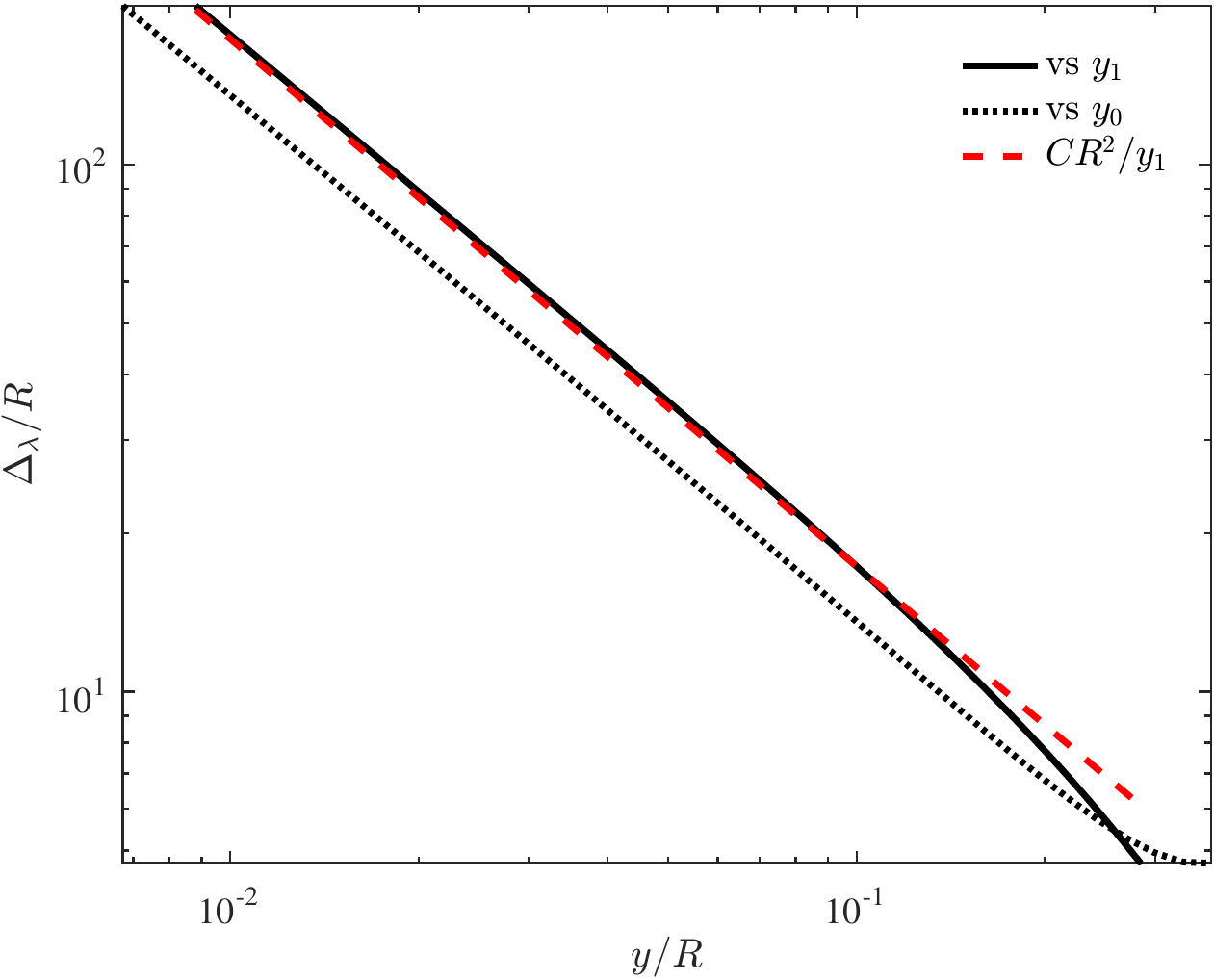}
    \caption{Net particle displacement~$\Delta_\lambda$ as a function of the the initial ($\yc_0$, dotted) and final ($\yc_1$, solid) distances from the swimming axis.  The particle starts far ahead of the swimmer and ends far behind.  The dashed line is the asymptotic form~\eqref{eq:nearfieldasymdisp}, which agrees with the displacement as a function of the final position~$\yc_1$.}
    \label{fig:puller_smally0}
  \end{center}
\end{figure}

For large~$\lambda$, the largest displacement values will involve particles
that travel with the swimmer for a long distance, i.e., particles that stay
near the swimmer's body.  The displacement in the~$\yc$ direction is then
negligible.  The residence time near the swimmer's body is
\begin{equation}
  \Tres = \int_0^T\dint\time =
  \RR\int_0^\pi
  \frac{\!\dint\theta}{\tavg{\uc_\theta(\delta\rc(\theta),\theta,\cdot)}}.
\end{equation}
We insert into this the velocity~\eqref{eq:Uavg} to get
\begin{equation}
  \Tres =
  \RR\int_0^\pi
  \l(\tfrac32\frac{\Uc}{\RR}\,\delta\rc(\theta)\sin\theta\l\{1
  + \Wavg(\theta)\r\}\r)^{-1}\!\dint\theta
\end{equation}
and then use the streamline~\eqref{eq:rrtheta} to find the net
displacement
\begin{equation}
  \Delta_\lambda(\yc_1) = \Uc \Tres =
  \sqrt{\tfrac23}\,\frac{\RR^2}{\yc_1}\int_0^\pi
  \l\{1 + \Wavg(\theta)\r\}^{-1/2}\!\dint\theta.
  \label{eq:Xdisp}
\end{equation}
This is independent of~$\xc$ since we assume the swimmer moves a long enough distance so that the particle crawls along the full length of the body.  We can evaluate the integral~\eqref{eq:Xdisp} numerically to find
\begin{equation}
  \Delta_\lambda(\yc_1) = \Cnear\RR^2/\yc_1,
  \qquad
  \Cnear \approx 1.72919.
  \label{eq:nearfieldasymdisp}
\end{equation}
The corresponding coefficient for an isolated no-slip sphere is~$\sqrt{2/3}\,\pi
\approx 2.56510$, so the net particle displacement is about~$67\%$ of an
equivalent sphere.  This asymptotic expression is compared to numerical simulations in Fig.~\ref{fig:puller_smally0}, showing excellent agreement.  Note that this predicts very large displacements for small~$\yc$, but in practice these will be capped by the swimming path length~$\lambda$.

\section{Far-field asymptotics}
\label{sec:farfield}

As we zoom out from our swimmer and look in the far-field, the force singularities in the flow field of Section~\ref{subsec:flowField} cancel out, as we required for our neutrally buoyant swimmer.  The net velocity field in~\eqref{eq:fullFlowField} is then well approximated for~$\lVert\rv\rVert/\Asep \gg 1$ by a stresslet singularity (with a source term for mass conservation),
\begin{equation}
  \uvstress(\rv)=\tfrac34\l(1-\frac{3\xc^2}{\lVert\rv\rVert^2}\r)\frac{\RR^2\rv}{\lVert\rv\rVert^3},
  \label{eq:betastressapprox}
\end{equation}
with stresslet strength
\begin{equation}
  \beta(\time) =
  \l[\l(\tfrac52 - \tfrac32\RRAsep^2\r)\RRAsep^2 - \Asep/\RR\r]
  (1+\da/\Uc)\,(\rr/\RR)\,.
  \label{eq:beta}
\end{equation}
Recall that~$\Asep=\aa-\AA$ is the separation between the flagellar Stokeslet and the center of the swimmer's body, and~$c_\Asep=\RR/\Asep$.  For the remainder of this section we will set~$\RR=1$ for expediency.

In the lab frame, particles obey~$\dot{\rvp}=\Uc\beta(\time)\,\uvstress(\rvp-\Uc\time\,\xuv)$.  Any time-dependence on the oscillatory positions and strengths of the original Stokeslets is absorbed by the stresslet strength~$\beta(\time)$ in~\eqref{eq:betastressapprox}.  The stresslet strength~$\beta(\time)$ has a Fourier series derived from~\eqref{eq:beta} which we analyze in the following two subsections.

\subsection{Displacement due to mean flow}
\label{sec:meanflowdisp}

In the lab frame, the stresslet starts at the origin and proceeds to move in the positive~$\xc$-direction with speed~$\Uc$. The mean flow from the swimmer is
\begin{equation}
  \uv(\rv,\time) = \Uc\betak{0}\uvstress(\rv - \Uc\time\,\xuv),
  \label{eq:uvmeanflow}
\end{equation}
where $\betak{0}=\tavg{\beta}\approx 5.2$ and~$\uvstress$ is defined in~\eqref{eq:betastressapprox}.  Let $\delta\rvp(\time) = \rv(\time) - \rv_0$ be the particle's displacement from $\rv_0$. If the particle is moderately far from the swimmer, then $\delta\rvp$ remains small throughout the trajectory, and we can expand~\eqref{eq:uvmeanflow} to leading order in $\delta\rvp$ as
\begin{equation}
  \uv(\rv,\time) = \Uc\betak{0}\uvstress(\rv_0 - \Uc\time\,\xuv)
  + \Order{\lVert\delta\rvp\rVert}.
  \label{eq:ufar}
\end{equation}
At this order the particle feels a velocity field that depends solely on its initial position. We can then solve for the particle motion~\eqref{eq:dotrvp} by integrating~\eqref{eq:ufar} directly to obtain
\begin{subequations}
  \begin{align}
    \delta\xcp(\time)
    &=\tfrac34\betak{0}\frac{\Hyp^2(\sqrt{2}\xc_0,\yc_0)}{\Hyp^3(\xc_0,\yc_0)}
    -\tfrac34\betak{0}\frac{\Hyp^2(\sqrt{2}(\xc_0-\Uc\time),\yc_0)}{\Hyp^3(\xc_0-\Uc\time,\yc_0)}, \\
    \delta\ycp(\time)
    &=\tfrac34\betak{0}\frac{\xc_0\yc_0}{\Hyp^3(\xc_0,\yc_0)}
    -\tfrac34\betak{0}\frac{(\xc_0-\Uc\time)\yc_0}{\Hyp^3(\xc_0-\Uc\time,\yc_0)},
  \end{align}
  \label{eq:dispfar}
\end{subequations}
valid to leading order in~$\delta\rvp$. Here the distance function is
\begin{equation}
  \Hyp(\xc,\yc)\ldef\sqrt{\xc^2+\yc^2}.
\end{equation}

\begin{figure}
  \begin{center}
    \begin{subfigure}[b]{.49\textwidth}
      \includegraphics[height=.28\textheight]{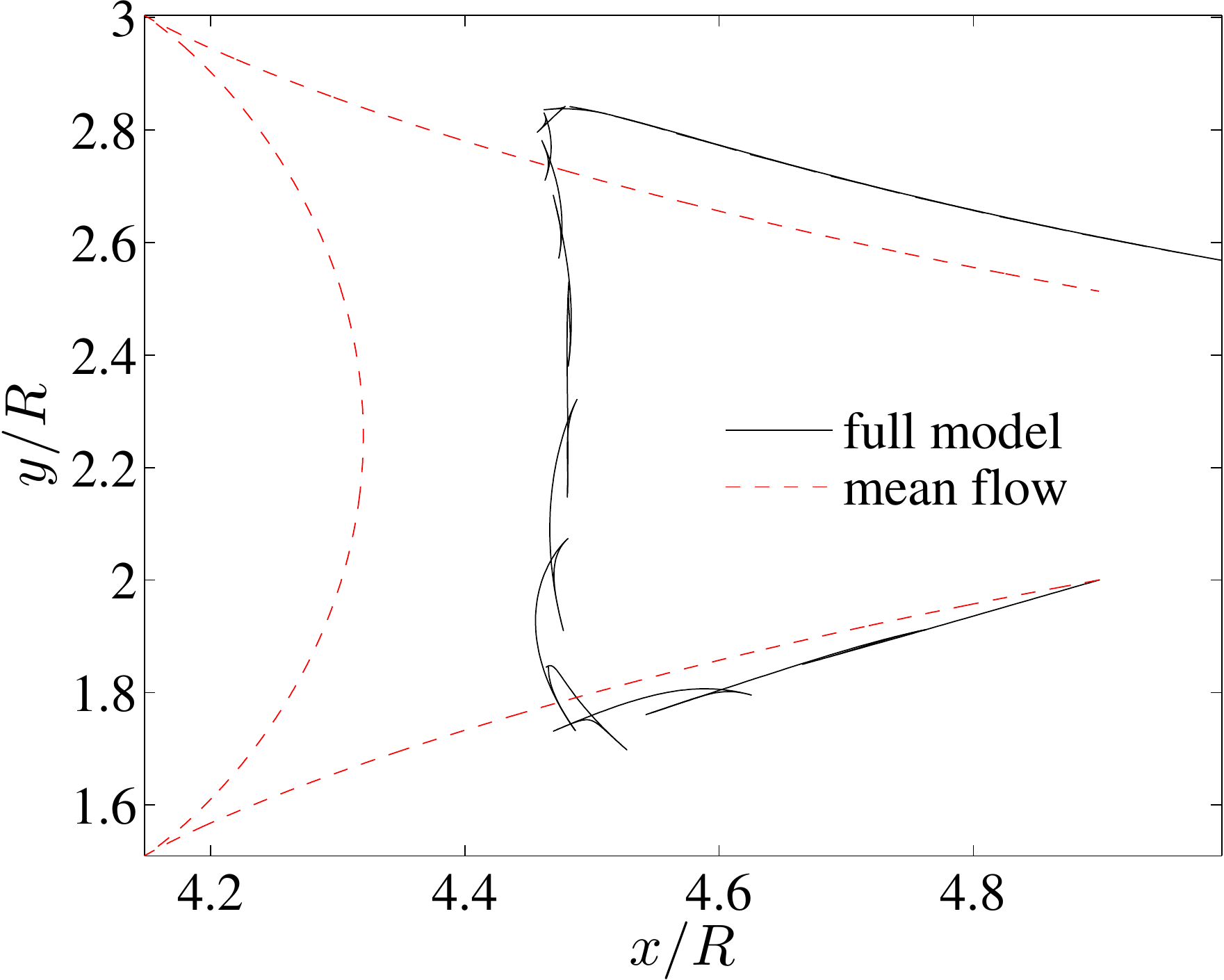}
      \caption{}
      \label{fig:farfield_close}
    \end{subfigure}
    \begin{subfigure}[b]{.49\textwidth}
      \includegraphics[height=.28\textheight]{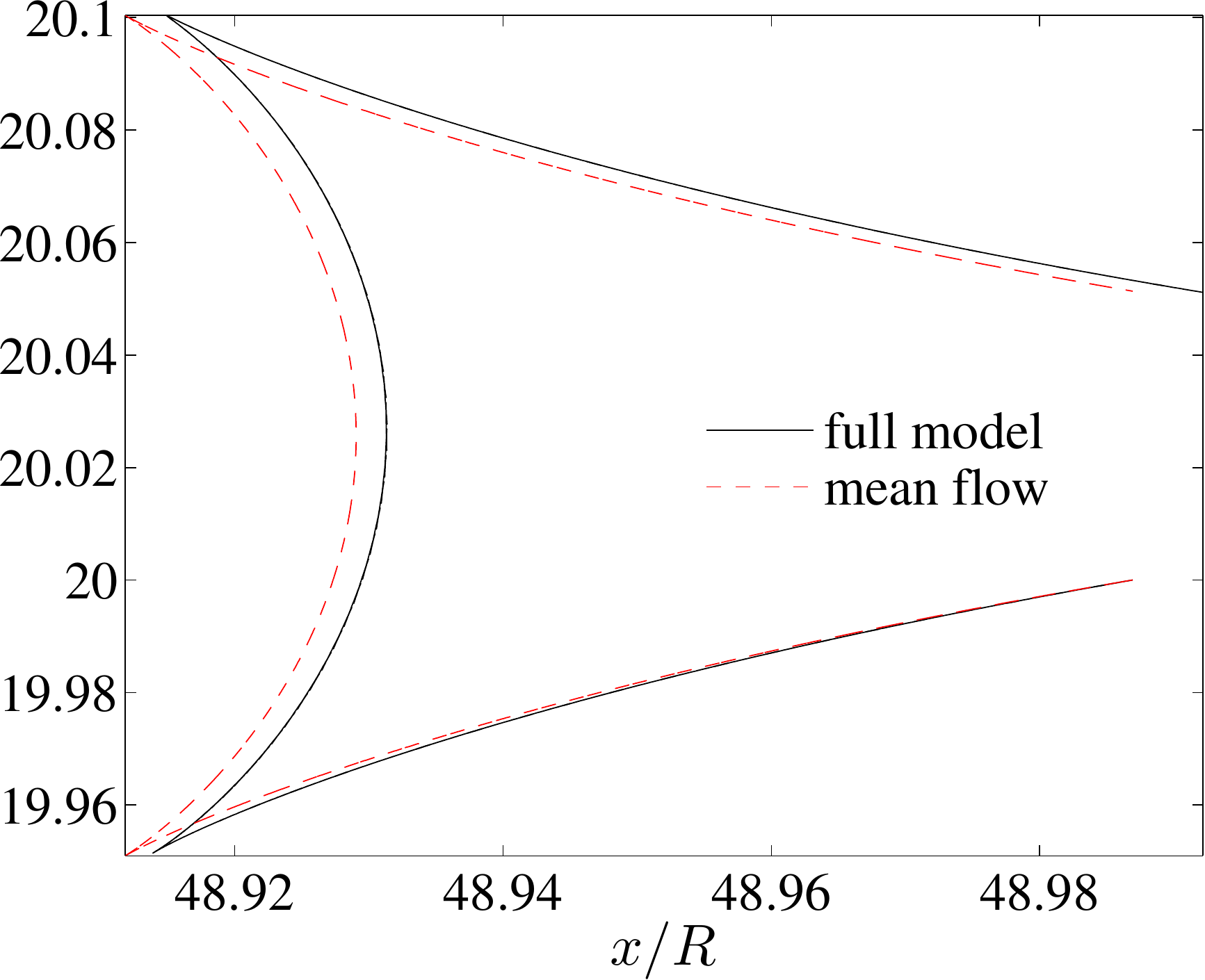}
      \caption{}
      \label{fig:farfield_far}
    \end{subfigure}
  \end{center}
  \caption{Particle paths (a) near the swimmer and (b) far from the swimmer. Paths caused by the full model (solid lines) from Section~\ref{subsec:flowField}, and the far-field approximation of the mean flow (dashed lines) from Section~\ref{sec:meanflowdisp}. Better agreement is seen for particle paths further from the swimmer.}
  \label{fig:farfield}
\end{figure}

Both coordinates achieve extrema at~$\Uc\time=\xc_0 \pm \tfrac{1}{\sqrt{2}}\yc_0$, and~$\delta\xcp(\time)$ has an additional extremum at~$\Uc\time=\xc_0$. The fact that both coordinates achieve extrema at the same time is reflected by the two `cusps' visible in Fig.~\ref{fig:farfield_far}. The coordinates of the two cusps are
\begin{equation}
  \delta\xc_{\mathrm{cusp}}=
  -\sqrt{\tfrac23}\frac{\betak{0}}{\lvert\yc_0\rvert}
  +\tfrac34\betak{0}\frac{\Hyp^2(\sqrt{2}\xc_0,\yc_0)}{\Hyp^3(\xc_0,\yc_0)},
  \quad
  \delta\yc_{\mathrm{cusp}}=
  \pm\tfrac{1}{2\sqrt{3}}\frac{\betak{0}}{\lvert\yc_0\rvert}
  +\tfrac34\frac{\betak{0}\,\xc_0\,\yc_0}{\Hyp^3(\xc_0,\yc_0)}.
  \label{eq:cusps}
\end{equation}
Examining Fig.~\ref{fig:farfield_far} and using the location of the cusps~\eqref{eq:cusps}, we find that the maximum displacements are bounded as
\begin{equation}
  \lvert\delta\xcp(\time)\rvert\leq\sqrt{\tfrac23}{\betak{0}}/{\lvert\yc_0\rvert},
  \quad\quad
  \lvert\delta\ycp(\time)\rvert\leq\tfrac{1}{\sqrt{3}}{\betak{0}}/{\lvert\yc_0\rvert}.
\end{equation}
The total net displacement after a time~$\time=\lambda/\Uc$ is
\begin{equation}
  \Deltafar_\lambda(\xc_0,\yc_0) =
  \Hyp(\delta\xcp(\lambda/\Uc),\delta\ycp(\lambda/\Uc))
  \le {\betak{0}}/{\lvert\yc_0\rvert}.
  \label{eq:stressDrift}
\end{equation}

\subsection{Displacement due to time-dependent flow}
\label{sec:timedepflowdisp}

In Section~\ref{sec:meanflowdisp} we ignored the time-dependence of~$\beta$ and focused on the mean flow. For small particle displacements the expansion~\eqref{eq:ufar} holds, and the velocity field measured at the particle only depends in the initial position of the particle relative to the swimmer, at leading order. This means we can consider the Fourier terms of~$\beta$ separately.
The Fourier series expansion of~$\beta(\time)$ is
\begin{equation}
  \beta(\time) = \sum_{m=-\infty}^\infty\betak{m}\ee^{\imi m\Omega\time}.
  \label{eq:infBetaFourier}
\end{equation}
Recall~$\betak{0}=\tavg{\beta}$ is the mean-flow portion described in Section~\ref{sec:meanflowdisp}. From~\eqref{eq:ufar}, the contribution to the displacement for a given frequency~$m\Omega$ will lead to the integral
\begin{equation}
  \delta\rvk{m}(\time)
  = \Uc\betak{m}
  \int_0^\time\uvstress(\xc_0 - \Uc s,\yc_0)\,\ee^{\imi m\Omega s}\dint s
  \label{eq:timedepint}
\end{equation}
where~$\uvstress$ is defined in~\eqref{eq:betastressapprox}.

At high frequencies we expect little contribution from the oscillating
part. Indeed, integrating~\eqref{eq:timedepint} by parts gives
\begin{equation}
  \int_0^t\uvstress(\xc_0 - \Uc s,\yc_0)\,\ee^{\imi m\Omega s}\dint s
  = \frac{1}{(\imi m\Omega)}
  \l[\uvstress(\xc_0 - \Uc s,\yc_0)\,\ee^{\imi m\Omega s}\r]_0^t
  +
  \Order{\Omega^{-2}}
  \label{eq:asympseries}
\end{equation}
for~$m\geq 1$.  From~\eqref{eq:timedepint}, the ratio of the contribution of
the~$\betak{m}$ term to the averaged flow ($m=0$) is roughly
\begin{equation}
  \frac{\lVert\delta\rvk{m}(\time)\rVert}{\lVert\delta\rvk{0}(\time)\rVert}
  \sim
  \frac{\Uc}{\Omega}\frac{\lvert\betak{m}\rvert}{\lvert\betak{0}\rvert}
  \frac{1}{\Hyp(\xc_0 - \Uc\time,\yc_0)}.
  \label{eq:betambeta0}
\end{equation}
We see that the ratio of displacements becomes smaller not only as~$\Omega$
becomes larger, but also as the distance~$\Hyp(\xc_0 - \Uc\time,\yc_0)$ is
made larger.  This is significant: it means that the time dependence has a
smaller relative impact on faraway particles than on nearby ones, in addition
to the averaging effect due to large~$\Omega$.  Hence, in the far-field, where
the stresslet approximation is valid, the time dependence of the swimmer can
be safely neglected.


\section{Statistics of particle displacements}
\label{sec:stats}

So far we have considered the displacements due to a single swimming organism.   However, there are several experiments such as \citet{Leptos2009}, \citet{Kurtuldu2011} and \citet{Jepson2013}, where microparticles are tracked in a bath of swimming organisms.  To model these experiments, we have to average over the random orientations of swimmers in an appropriate manner.  We follow here the procedure of \cite{Lin2011,Thiffeault2010b} for finding the effective diffusivity, and of \cite{Thiffeault2015} for obtaining the full probability distribution function of particle displacements.  We find that including a no-slip body, as in our present model, `lifts' the tails of the distribution by making large displacements more common.

\subsection{Effective diffusivity}
\label{sec:effDiff}

\begin{figure}
  \begin{center}
    \includegraphics[width=.6\textwidth]{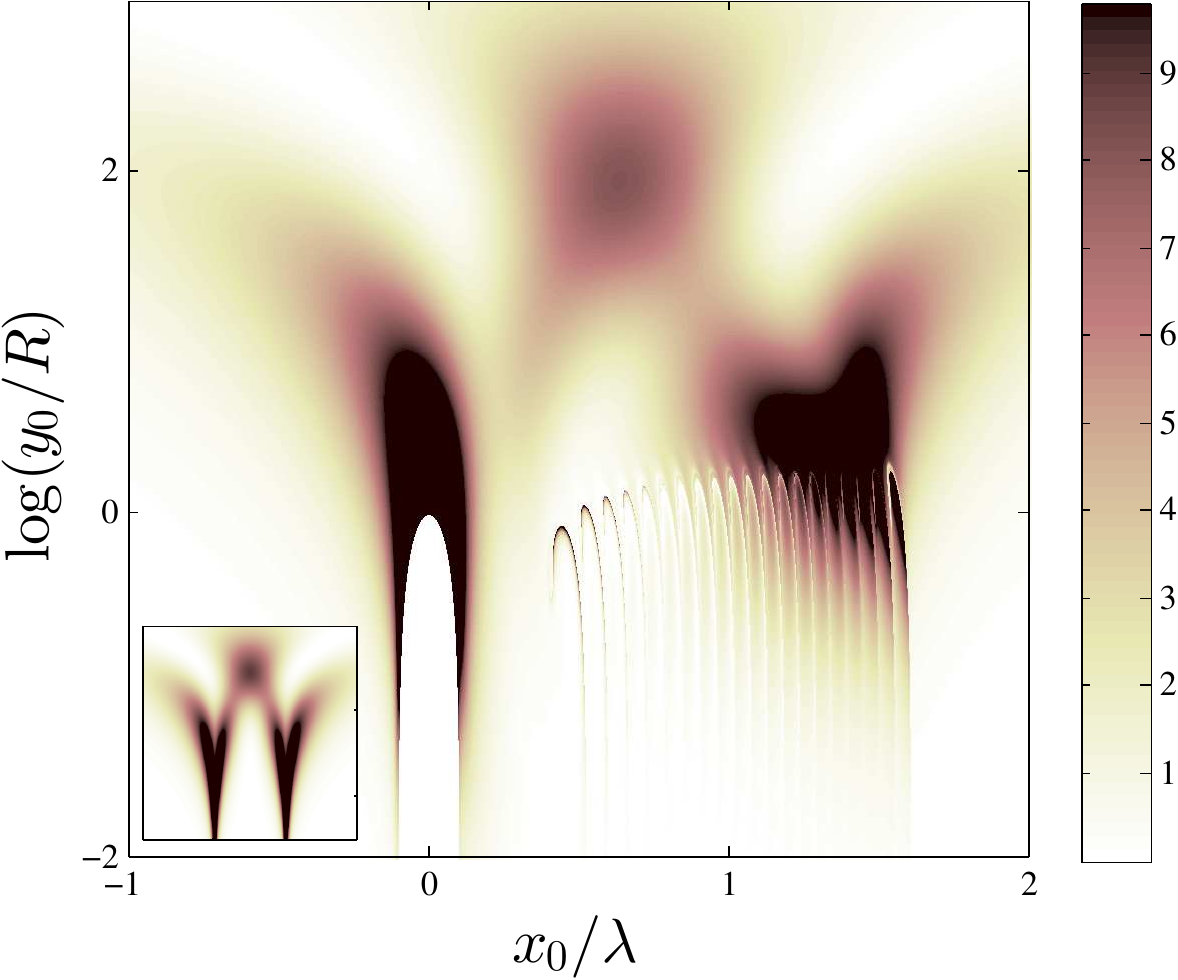}
    \caption{Integrand of the enhanced diffusivity integral~\eqref{eq:Denhancedintaxi} with~$\lambda\approx 9.8\lengthscale$. The inset is the steady stresslet approximation, with~$\betak{0}\approx 5.2\lengthscale^2$.}
    \label{fig:secondMomentIntegrand}
  \end{center}
\end{figure}

\begin{figure}
  \begin{center}
    \begin{subfigure}[b]{.46\textwidth}
      \includegraphics[height=.25\textheight]{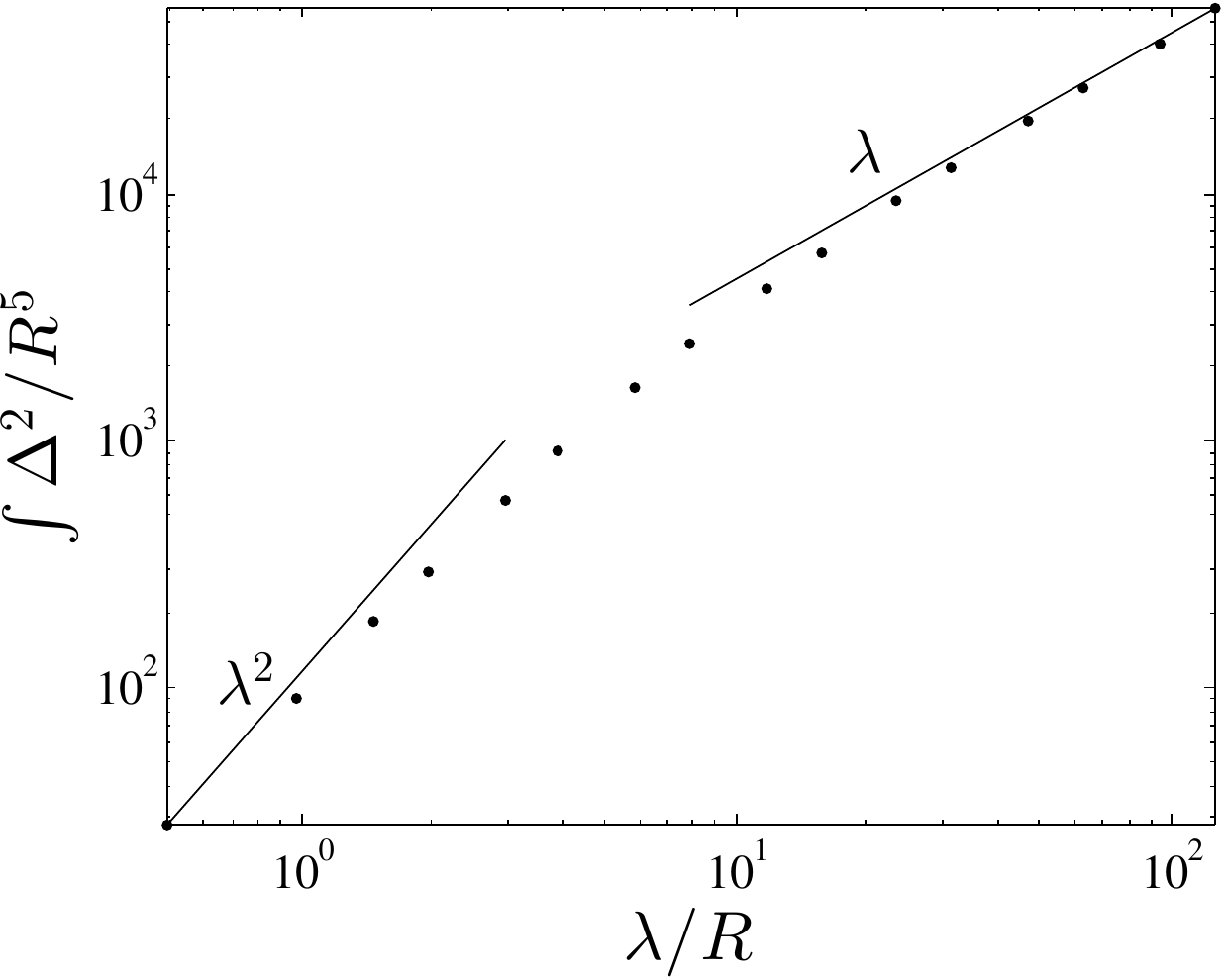}
    \end{subfigure}
    \hspace{.25em}
    \begin{subfigure}[b]{.46\textwidth}
      \includegraphics[height=.25\textheight]{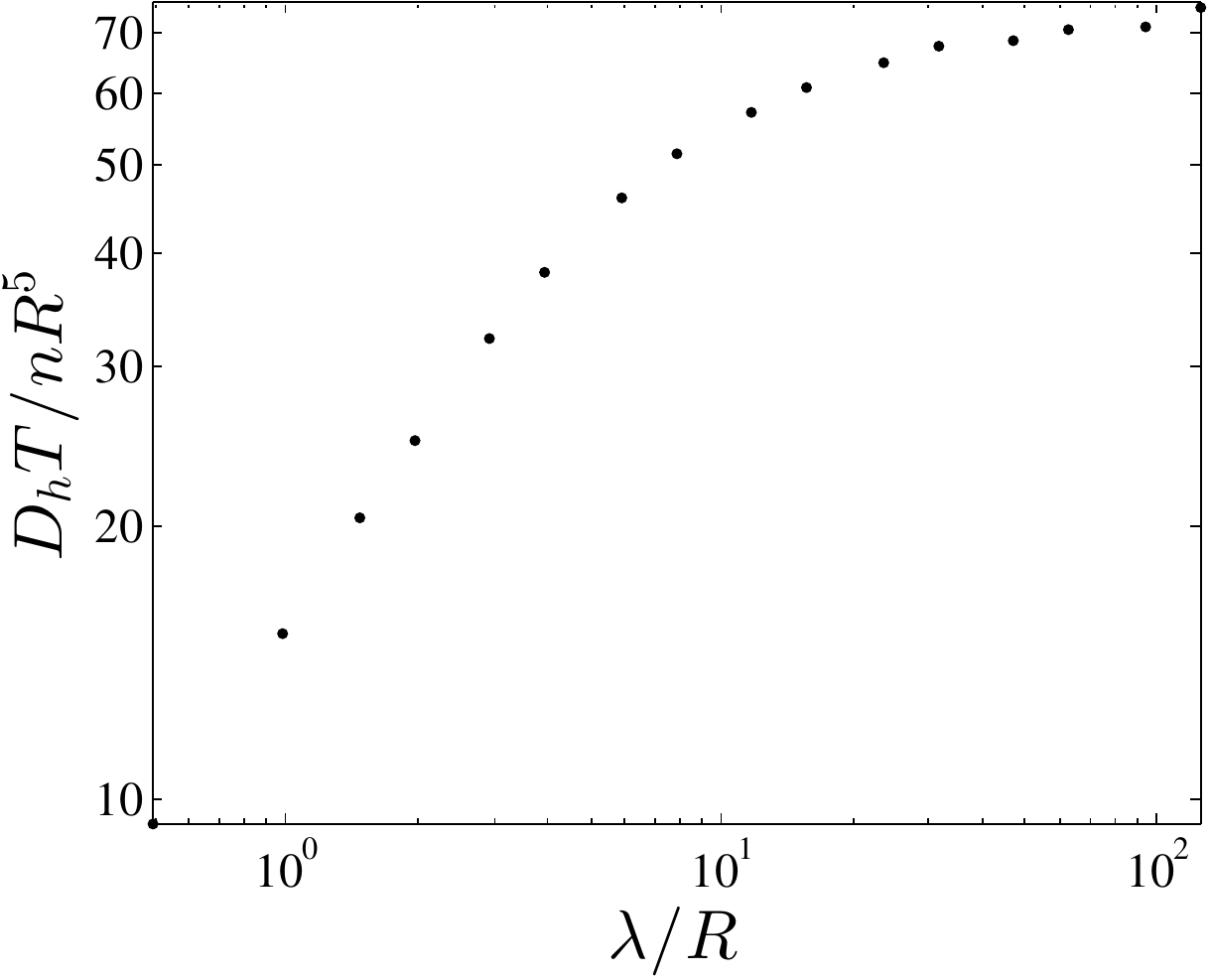}
    \end{subfigure}
    \caption{(a) Dimensionless values of the second moment of particle displacements and (b) effective diffusivity for varying path lengths of swimmers.}
    \label{fig:logdispintTspan}
  \end{center}
\end{figure}

At low  swimmer volume fractions, the effective diffusivity~($\Deffective$) separates into a thermal diffusivity~($\Dthermal$) and an \enhanced~diffusivity~($\Denhanced$)~\cite{Jepson2013,Mino2011,Pushkin2013b,Underhill2008}. The \enhanced~diffusivity measures how the swimmers affect their environment in the absence of thermal noise (which our numerics and asymptotics also neglect), and is related to the second moment of particle displacements via
\begin{equation}
  \Denhanced = \frac{\ndens\Uc}{6\lambda}\int_{\mathbb{R}^3}
  \Delta_\lambda^2(\rv_0)\dint^3\rv_0,
  \label{eq:Denhancedint}
\end{equation}
where~$\ndens$ is the number density of the swimmers \cite{Thiffeault2010b,Lin2011,Pushkin2013b}.  Here the integral is over all possible initial positions of a fluid particle with respect to the swimmer, assuming an infinite domain (and convergence of the integral --- see~\cite{Thiffeault2015}).

The axial symmetry of our swimmer simplifies~\eqref{eq:Denhancedint} to
\begin{equation}
  \Denhanced
  =\tfrac13\pi\Uc \ndens\lambda^{-1}\int_{\mathbb{R}^2}\yc_0^2\,\Delta_\lambda^2(\xc_0,\yc_0)\dint\log(\yc_0/\RR)\dint\xc_0,
  \label{eq:Denhancedintaxi}
\end{equation}
where we use~$\log(\yc_0/\RR)$ as the integration variable to emphasize small-$\yc_0$ values, for which the largest displacements occur.  We used the axial symmetry to treat~$\yc_0$ like a perpendicular distance from the~$\xc_0$-axis (the swimming axis).  A sample integrand for~$\lambda\approx 9.8\lengthscale$ is plotted in Fig.~\ref{fig:secondMomentIntegrand}.  This is closely related to the particle displacement plot Fig.~\ref{fig:dispint01}, with the addition of the log scaling and the axial symmetry weight~$\yc_0$ which measures the `rarity' of close encounters~\cite{Lin2011}.  The inset in Fig.~\ref{fig:secondMomentIntegrand} shows the far-field stresslet form, which is not valid near the swimmer.  The largest displacements have been smeared by the time-dependence, and are now asymmetric with respect to the start and end of the swimming path.  The largest displacements are associated with particles dragged along the swimmer's no-slip body.  However, these are not the dominant contribution to the integral~\eqref{eq:Denhancedintaxi}, because of the~$\yc_0$ weight.  The largest displacements are too rare to significantly affect the enhanced diffusivity.

Values of the integral~\eqref{eq:Denhancedintaxi} are plotted for varying path length in Fig.~\ref{fig:logdispintTspan}(a), with~$\time = \lambda/\Uc$.  We observe a roughly `ballistic' scaling ($\lambda^2$) for short swimming times, and a diffusive scaling ($\lambda$) for longer times.  This is consistent with the observations in \cite{Wu2000,Underhill2008,Leptos2009,Thiffeault2015}: for short times particles move linearly in time, and so the squared displacement is quadratic with~$\lambda$.  (In some of these publications the exponent seems smaller than ballistic, which could be because the data is already turning over to the diffusive regime, or because of molecular diffusion.)  For longer times particles are left behind and undergo a finite displacement, but the number of particles displaced grows linearly with~$\lambda$~\cite{Thiffeault2015}.  In the far-field the displacements due to a stresslet singularity also leads to linear dependence on~$\lambda$, as described by~\citet{Pushkin2013b} and~\citet{Thiffeault2015}.

In Figure.~\ref{fig:logdispintTspan}(b) we see that the effective diffusivity eventually saturates with path length~$\lambda$, reaching an asymptotic value of about~$70$ in dimensionless units.  For comparison, if we use only the far-field averaged stresslet value, we find a value of about~$60$.  The increase in the enhanced diffusivity due to time dependence and modeling of the near field is thus significant but not large.  This is consistent with the observation that the integral in~\eqref{eq:Denhancedintaxi} is dominated by particles that are a few radii away from the swimmer~\cite{Lin2011}, where the stresslet approximation will start to apply, and the heavy suppression of the time dependence at those distances as reflected by~\eqref{eq:betambeta0}.


\subsection{Distribution of particle displacements}
\label{sec:pdf}

In \citet{Thiffeault2015}, the experimental results of \citet{Leptos2009} were well-explained by examining the drift function due to a model organism, called a squirmer.  Squirmers were introduced by \citet{Lighthill1952} and \citet{Blake1971}; they consist of a sphere in Stokes flow with an imposed tangential velocity.  The force-free condition is imposed to determine the swimming velocity.  The far-field form of the velocity field is thus a stresslet, as required for a neutrally-buoyant microswimmer.  The imposed velocity at the surface of the squirmer leads to lessened largest particle displacements compared to the model presented here, since particles are not dragged along by the squirmer.

The experimental distributions of \citet{Leptos2009} were well-fitted at different volume fractions by steady squirmers with a stresslet strength~$\beta=0.5$.  However, it was observed that the fit was worst in the tails of the distribution, corresponding to the largest particle displacements.  The hypothesis in modeling a more realistic swimmer with a no-slip body was that this would lead to fatter tails while leaving the center of the distribution mostly unchanged, since the center depends mostly on far-field (stresslet) effects.

\begin{figure}
  \begin{center}
    \includegraphics[width=.32\textwidth]{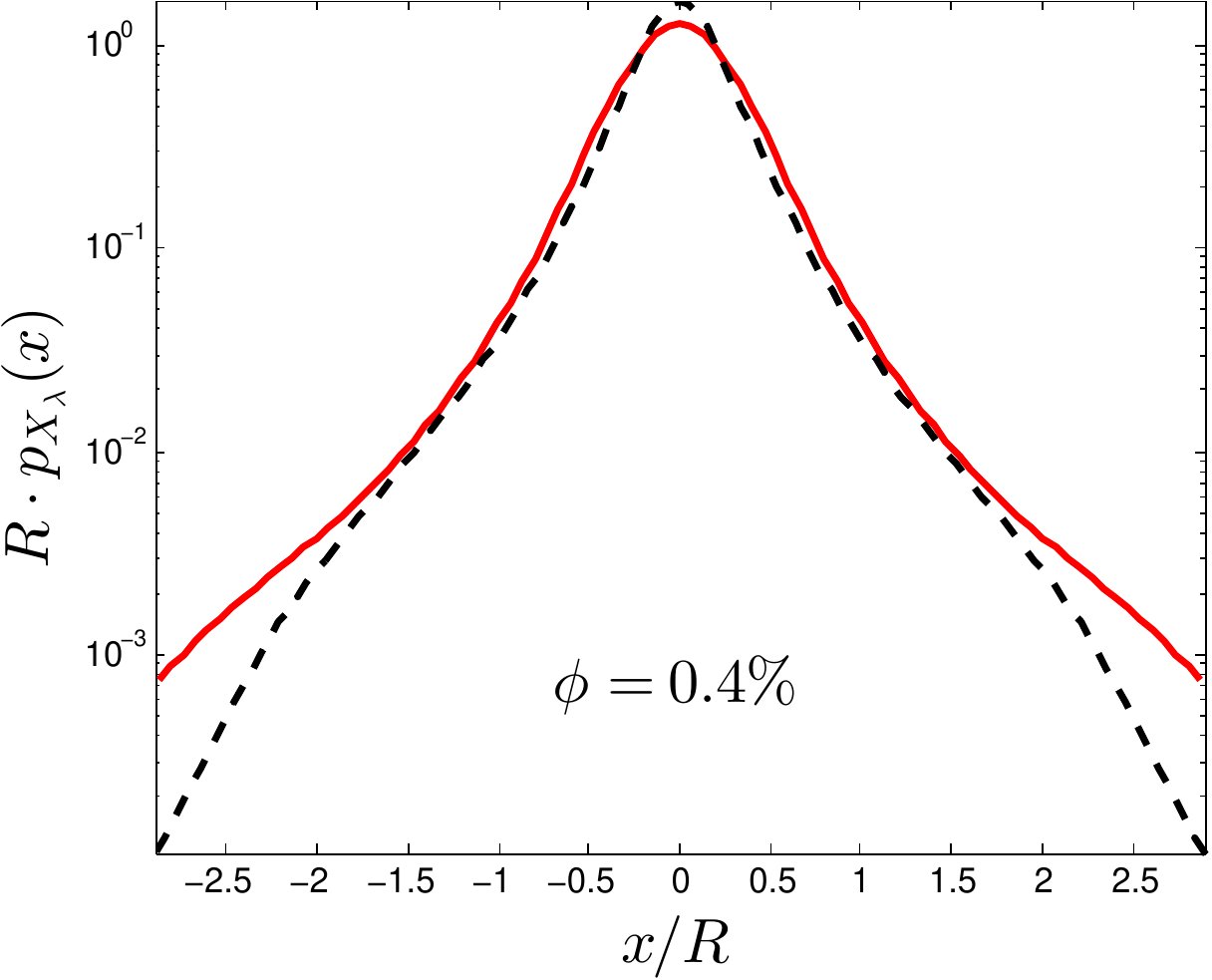}
    \includegraphics[width=.32\textwidth]{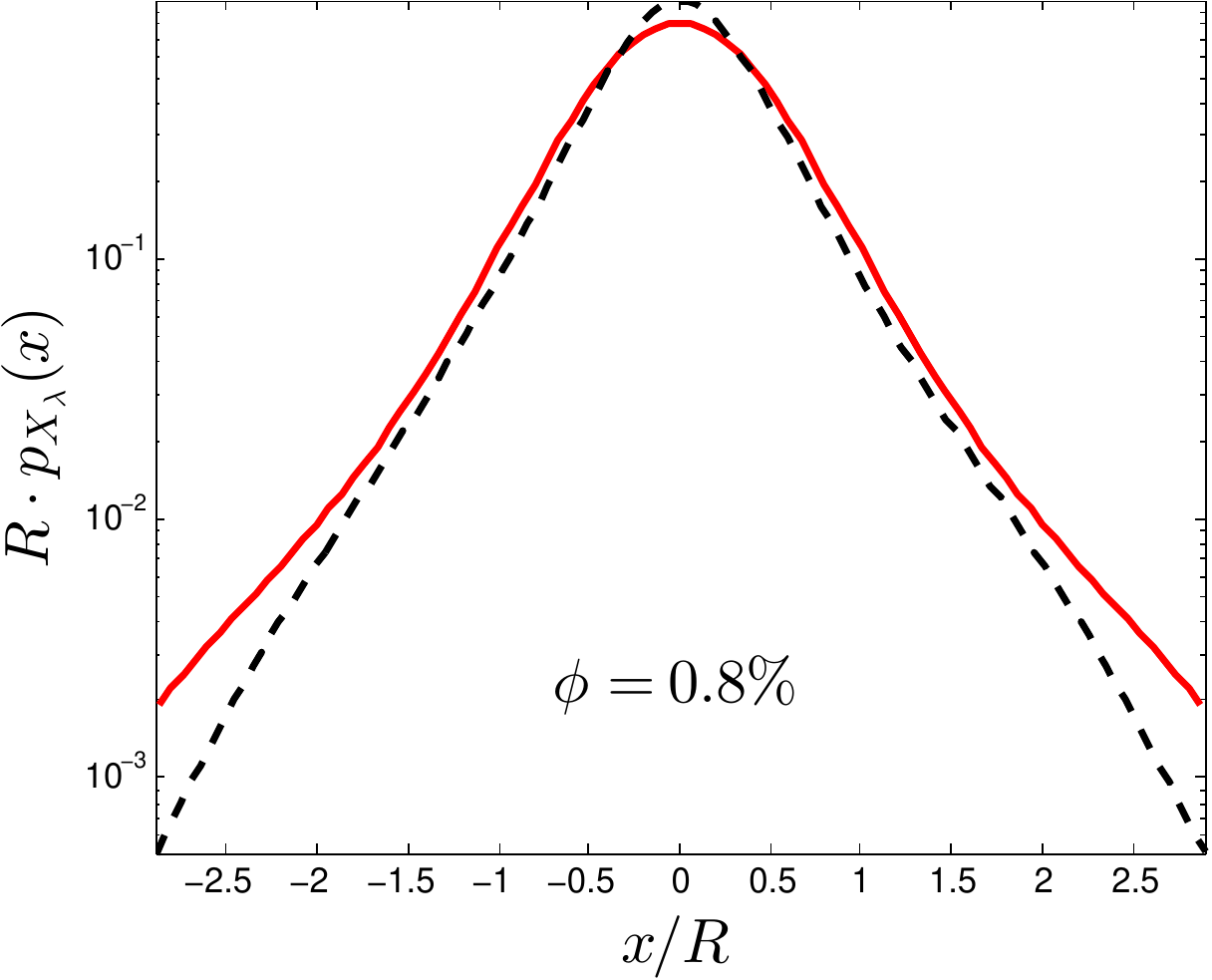}
    \includegraphics[width=.32\textwidth]{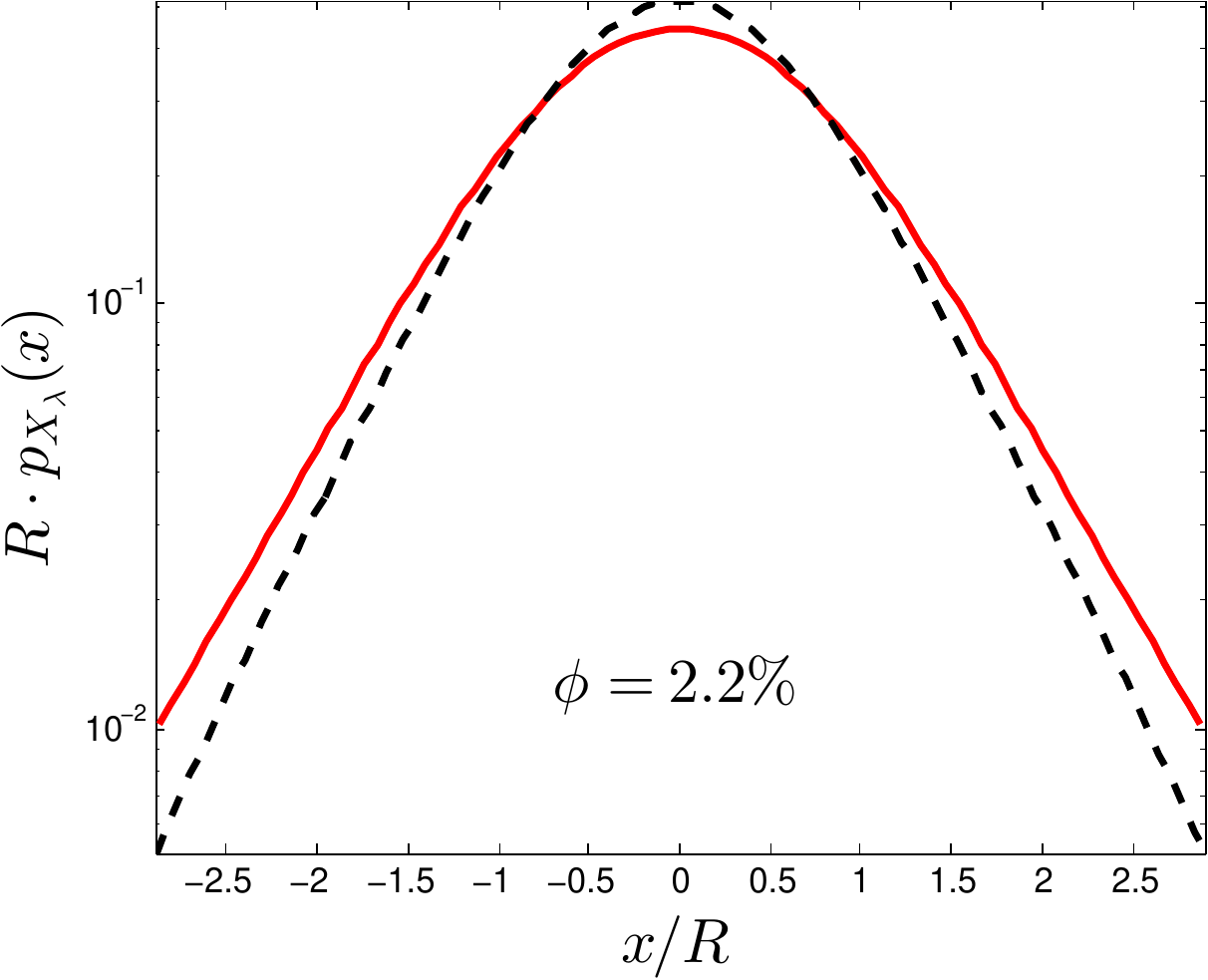}
    \caption{The probability distribution function for varying swimmer volume fractions for steady squirmers (dashed) and the time-dependent model in this paper (solid). The average stresslet strength~$\beta$ is the same in both models.   The swimmers move a net distance of~$\lambda/\RR\approx3$, which is comparable to the experiments of \citet{Leptos2009}.}
    \label{fig:compareModels}
  \end{center}
\end{figure}

In Fig.~\ref{fig:compareModels} we plot the probability distribution functions for a few volume fractions and compare our model to the steady squirmer for~$\beta=\tavg{\beta}=5.2$, the mean stresslet strength for our time-dependent swimmer.  As expected, the tails of the distribution are somewhat fatter in our time-dependent model with a no-slip sphere.  This improves the match to the data of \citet{Leptos2009}, though we did not directly compare to their data since the values of~$\beta$ required leads to somewhat unrealistic parameters in our model, such as the flagellum entering the body.  This can be explained by the fact that \emph{C.\ reinhardtii} has two flagella that can move to the sides of the body, whereas our model exploits axial symmetry to maintain its simplicity.  Note also that in computing the distributions for the squirmer in Fig.~\ref{fig:compareModels} we omitted particles in the `atmosphere' (trapped recirculation region) present at these values of~$\beta$, as described in \cite{Lin2011}, since such an atmosphere is absent from the time-dependent model.  Figure~\ref{fig:compareModels} also highlights the convergence to a more Gaussian form as the volume fraction is increased, though the distribution is still far from Gaussian \cite{Thiffeault2015}.

\section{Conclusion}
\label{sec:conc}

In this paper we have modeled a microswimmer as a no-slip sphere for the swimmer's body and a time-dependent point force (Stokeslet) for its flagella.  The model is closer in its dynamical appearance to realistic organisms, such as \emph{C.\ reinhardtii}.  We then compute the time-dependent drift of particles advected by the swimmer.  Near the swimmer, we see the stretching and folding action typical of chaotic systems (Fig.~\ref{fig:cloud_drift}).  We did not investigate this fully, though it would be interesting to examine the small-scale mixing due to microswimmers using the tools from transient chaos in open flows~\cite{Tel1996,Pentek1999}.

The drift function, which describes the displacements of fluid particles as the swimmer moves a finite distance, is an interesting object of study in its own right.  However, for a time-dependent swimmer we must rely mostly on numerical simulations, as we have done here.  The asymptotics of the drift function for the largest displacements (near the swimmer) and the smallest (far away from the swimmer) are also important to understand when examining particle statistics, since these depend on integrals of the drift function over all space.  We found that for the largest displacements the drift function exhibits the~$1/\yc_0$ singularity typical of a no-slip sphere, which corresponds to particles hugging the swimmer's body.  However, the drift distance is reduced when compared to an isolated sphere, since the flagellar Stokeslet pushes particles along the body.  We were able to obtain a rough estimate for the drift near the body by a suitable averaging over the fast swimming stroke period.

Far from the swimmer, we expect the time dependence to be damped.  We showed this explicitly by using the standard method of repeated integration by parts for developing an asymptotic expansion in a fast variable.  An important outcome is that the time dependence is damped in two ways: it is damped because it is fast, but also it decreases inversely with distance.  Thus, the time dependence is unimportant in many applications that only depend on particle displacements a few radii away from the swimmer.

One application in which the large time-dependent displacements are important
is to the statistics of particle displacements.  In previous work~\cite{Thiffeault2015} the experimental distributions of particle displacements of \citet{Leptos2009} were well-matched by a steady squirmer model.  However, the non-Gaussian tails, which are associated with large displacements, were found to be somewhat below the experiments, indicating that the steady model underestimated the probability of large displacements.  We find here that the combination of time dependence and the presence of a no-slip boundary raise these tails while leaving the center of the distribution relatively unchanged.  We were not able to match to the experimental distributions themselves: even though our model used parameters close to \emph{C.\ reinhardtii}, the axial symmetry we used makes matching the mean stresslet strength of that organism very difficult (it would require the flagellar singularity to enter the body, which is unrealistic).  Obviously, a better model would be to use two flagella such as in \cite{Friedrich2012}, but breaking axial symmetry makes the necessary volume integrals much harder to evaluate.  In addition, there are enough additional parameters that simply matching the experimental distribution with this new model would not be very convincing.  (The fit in \cite{Thiffeault2015} required the adjustment of only one parameter, the mean stresslet strength.)  It may be possible in future experiments to measure the drift function directly, which would help discriminate between models.

\begin{acknowledgments}
  The authors are grateful to Mike Graham and Saverio Spagnolie for helpful discussions. The computations were made possible by the UW-Madison Center for High Throughput Computing.  The research was supported by NSF Grants DMS-1109315 and DMS-1147523.
\end{acknowledgments}

\bibliography{../../bib/journals_abbrev,../../bib/articles}

\end{document}

%% file: figs_tikz/chlamy.tex
\begin{tikzpicture}[scale=1.25,rotate=270]
  \newcommand\majrad{1.2}
  \newcommand\minrad{1}
  \newcommand\brad{.75} 
  \shadedraw[ball color=green] (0,0) ellipse (\minrad*\brad cm and \majrad*\brad cm);
  \draw[ultra thick,-latex] ({-3.5*\brad*\minrad*sin(50)},{3.5*\brad*\majrad*cos(50)})
    arc (140:180:{.5*\brad} and {\brad});
  \draw[ultra thick,-latex] ({3.5*\brad*\minrad*sin(50)},{3.5*\brad*\majrad*cos(50)})
    arc (40:0:{.5*\brad} and {\brad});

  \draw[ultra thick,gray!75] plot [smooth, tension=.5] coordinates {({-\brad*\minrad*sin(10)},{\brad*\majrad*cos(10)}) ({-1.2*\brad*\minrad*sin(10)},{1.2*\brad*\majrad*cos(10)}) ({-1.4*\brad*\minrad*sin(15)},{1.4*\brad*\majrad*cos(15)}) ({-1.5*\brad*\minrad*sin(45)},{1.5*\brad*\majrad*cos(45)}) ({-3.3*\brad*\minrad*sin(75)},{3.3*\brad*\majrad*cos(75)}) };
  \draw[ultra thick,gray!75] plot [smooth, tension=.5] coordinates {({\brad*\minrad*sin(10)},{\brad*\majrad*cos(10)}) ({1.2*\brad*\minrad*sin(10)},{1.2*\brad*\majrad*cos(10)}) ({1.4*\brad*\minrad*sin(15)},{1.4*\brad*\majrad*cos(15)}) ({1.5*\brad*\minrad*sin(45)},{1.5*\brad*\majrad*cos(45)}) ({3.3*\brad*\minrad*sin(75)},{3.3*\brad*\majrad*cos(75)}) };

  \draw[ultra thick,gray!75] plot [smooth, tension=.5] coordinates {({-\brad*\minrad*sin(10)},{\brad*\majrad*cos(10)}) ({-1.3*\brad*\minrad*sin(10)},{1.3*\brad*\majrad*cos(10)}) ({-1.45*\brad*\minrad*sin(15)},{1.45*\brad*\majrad*cos(15)}) ({-1.35*\brad*\minrad*sin(45)},{1.35*\brad*\majrad*cos(45)}) ({-1.8*\brad*\minrad*sin(100)},{1.8*\brad*\majrad*cos(100)}) ({-2.5*\brad*\minrad*sin(110)},{2.5*\brad*\majrad*cos(110)}) };
  \draw[ultra thick,gray!75] plot [smooth, tension=.5] coordinates {({\brad*\minrad*sin(10)},{\brad*\majrad*cos(10)}) ({1.3*\brad*\minrad*sin(10)},{1.3*\brad*\majrad*cos(10)}) ({1.45*\brad*\minrad*sin(15)},{1.45*\brad*\majrad*cos(15)}) ({1.35*\brad*\minrad*sin(45)},{1.35*\brad*\majrad*cos(45)}) ({1.8*\brad*\minrad*sin(100)},{1.8*\brad*\majrad*cos(100)}) ({2.5*\brad*\minrad*sin(110)},{2.5*\brad*\majrad*cos(110)}) };

  \draw[ultra thick,gray!75] plot [smooth, tension=.5] coordinates {({-\brad*\minrad*sin(10)},{\brad*\majrad*cos(10)}) ({-2*\brad*\minrad*sin(5)},{2*\brad*\majrad*cos(5)}) ({-2.35*\brad*\minrad*sin(6.5)},{2.35*\brad*\majrad*cos(6.5)}) ({-2.5*\brad*\minrad*sin(10)},{2.5*\brad*\majrad*cos(10)}) ({-2.5*\brad*\minrad*sin(15)},{2.5*\brad*\majrad*cos(15)}) ({-2.25*\brad*\minrad*sin(20)},{2.25*\brad*\majrad*cos(20)}) ({-1.5*\brad*\minrad*sin(35)},{1.5*\brad*\majrad*cos(35)})};
  \draw[ultra thick,gray!75] plot [smooth, tension=.5] coordinates {({\brad*\minrad*sin(10)},{\brad*\majrad*cos(10)}) ({2*\brad*\minrad*sin(5)},{2*\brad*\majrad*cos(5)}) ({2.35*\brad*\minrad*sin(6.5)},{2.35*\brad*\majrad*cos(6.5)}) ({2.5*\brad*\minrad*sin(10)},{2.5*\brad*\majrad*cos(10)}) ({2.5*\brad*\minrad*sin(15)},{2.5*\brad*\majrad*cos(15)}) ({2.25*\brad*\minrad*sin(20)},{2.25*\brad*\majrad*cos(20)}) ({1.5*\brad*\minrad*sin(35)},{1.5*\brad*\majrad*cos(35)})};

  \draw[ultra thick] plot [smooth, tension=.1] coordinates {({-\brad*\minrad*sin(10)},{\brad*\majrad*cos(10)}) ({-1.2*\brad*\minrad*sin(10)},{1.2*\brad*\majrad*cos(10)}) ({-1.4*\brad*\minrad*sin(15)},{1.4*\brad*\majrad*cos(15)}) ({-4*\brad*\minrad*sin(50)},{4*\brad*\majrad*cos(50)}) };
  \draw[ultra thick] plot [smooth, tension=.1] coordinates {({\brad*\minrad*sin(10)},{\brad*\majrad*cos(10)}) ({1.2*\brad*\minrad*sin(10)},{1.2*\brad*\majrad*cos(10)}) ({1.4*\brad*\minrad*sin(15)},{1.4*\brad*\majrad*cos(15)}) ({4*\brad*\minrad*sin(50)},{4*\brad*\majrad*cos(50)}) };

\end{tikzpicture}

%% file: figs_tikz/swimmer.tex
\begin{tikzpicture}[scale=1.25]
  \newcommand\pathradx{.75}
  \newcommand\pathx{2} 
  \newcommand\srad{.2} 
  \pgfmathsetmacro{\flagx}{\pathx}
  \shadedraw[ball color=green] (0,0) circle (.75);
  \draw[dashed,thick,-latex] 
    (\pathx-\pathradx,.25*\srad) -- (\pathx+\pathradx,.25*\srad);
  \draw[dashed,thick,-latex]
    (\pathx+\pathradx,-.25*\srad) -- (\pathx-\pathradx,-.25*\srad);
  \shadedraw[ball color=blue] (\flagx,0) circle (\srad);
  \draw (1.25*\flagx,-2*\srad) node {$(\aa,0,0)$};
  \draw (-.9,-.9) node {$(\AA,0,0)$};
  \draw[very thick,-latex] (.5,-1) -- (1.5,-1) node[right] {$\Uv$};
\end{tikzpicture}